\documentclass[aps,prb,twocolumn,unsortedaddress,superscriptaddress,showpacs]{revtex4-2}

\usepackage{color}
\usepackage{graphicx}
\usepackage{amsmath,amsfonts,amsthm} 
\usepackage{multirow}
\usepackage{amsmath,physics,siunitx} 
\usepackage{graphicx}
\usepackage[T1]{fontenc} 
\graphicspath{{./images/}}

\begin{document}

\title{Chirality-induced spin texture switching in twisted bilayer graphene}
\author{Kunihiro Yananose}
\affiliation{Center for Theoretical Physics, Department of Physics and Astronomy, Seoul National University, Seoul 08826, Republic of Korea}
\author{Giovanni Cantele}
\affiliation{Consiglio Nazionale delle Ricerche, Institute for Superconducting and Innovative Materials and Devices (CNR-SPIN), c/o Complesso di Monte S. Angelo, via Cinthia - 80126 - Napoli, Italy}
\author{Procolo Lucignano}
\affiliation{Universit\`a degli Studi di Napoli ``Federico II'',
	Dipartimento di Fisica ``Ettore Pancini'', Complesso di Monte S. Angelo, via Cinthia - 80126 - Napoli, Italy}
\author{Sang-Wook Cheong}
\affiliation{Rutgers Center for Emergent Materials and Department of Physics and Astronomy, Piscataway, NJ 08854, USA}
\author{Jaejun Yu}
\affiliation{Center for Theoretical Physics, Department of Physics and Astronomy, Seoul National University, Seoul 08826, Republic of Korea}
\author{Alessandro Stroppa}
\affiliation{Consiglio Nazionale delle Ricerche, Institute for Superconducting and Innovative Materials and Devices (CNR-SPIN), c/o Department of Physical and Chemical Sciences, University of L’Aquila, Via Vetoio, I-67100, Coppito, L’Aquila, Italy}

\date{\today}

\begin{abstract}
The interlayer van der Waals interaction in twisted bilayer graphene (tBLG) induces both in-plane and out-of-plane atomic displacements showing complex patterns that depend on the twist angle. 
In particular, for small twist angles, within each graphene layer, the relaxations give rise to a vortex-like displacement pattern which is known to affect the dispersion of the flat bands.
Here, we focus on yet another structural property, the chirality of the twisted bilayer. We perform first-principles calculations based on density functional theory to investigate the properties induced by twist chirality in both real and momentum space. In real space, we study the interplay between twist chirality and atomic relaxation patterns. In momentum space, we investigate the spin textures around the $K$ points of the Brillouin zone, showing that alternating vortex-like textures are correlated with the chirality of tBLG. Interestingly, the helicity of each vortex is inverted by changing the chirality while the different twist angles also modify the spin textures. We discuss the origin of the spin textures by calculating the layer weights and using plot regression models.
\end{abstract}

\maketitle

\section{Introduction}
Nowadays, two-dimensional (2D) materials represent a new exciting field in condensed matter physics and material science~\cite{khan_recent_2020, andrei_graphene_2020, barraza-lopez_colloquium_2021}.
The recent discovery of the so-called ``magic angle'' twisted bilayer graphene (mtBLG)~\cite{Cao:2018kn,Cao:2018ff} has attracted significant attention. When the twist angle between two stacked graphene sheets approaches the value of 1.08$^\circ$ which is called the first magic angle, the Fermi velocity becomes almost zero \cite{bistritzer_moire_2011,tarnopolsky_origin_2019}. 
In contrast to the linear band dispersion of the single-layer graphene, which implies a massless Dirac particle, the mtBLG shows an almost flat band dispersion with a tiny bandwidth that is of the order of $\sim 10$ meV as confirmed by tunneling spectroscopy experiments \cite{Kerelsky:2019aa,Xie:2019aa,Jiang:2019aa,Choi:2019aa}.
As the bands flatten at the Fermi level, the corresponding density of states tends to increase. Indeed, the kinetic energy of the flat-band electrons in mtBLG decreases so that Coulomb interactions among electrons are expected to play an essential role as like as in strongly correlated electron systems.
At $\theta\sim 1.08^\circ$, transport experiments show an intriguing phase diagram as a function of carrier concentration and other controlling parameters, displaying different superconducting domes as well as correlated insulating phases~\cite{Codecido:2019wy,Efetov:2019,Sharpe605,Yankowitz1059,Kennes2020,Choi:2019aa}.
The correlated insulating phase of mtBLG is indeed attributed to the enhanced electron-electron interactions within the flat bands~\cite{Sboychakov:2018tp,Rademaker:PRB2019}, although some authors are highlighting the relevance of the electron-phonon interactions~\cite{choi_strong_2018,Angeli:2019,Koshino_Son:2019,Lamparski_2020}.
The control of the flat bands has driven a significant interest on the theoretical study of tBLG showing that the twist angle in novel 2D structures can  be used as a further degree of freedom~\cite{RibeiroPalau:2018ki} for  implementing desired 
properties~\cite{Geim:2014hf, PhysRevMaterials.1.014002,Borriello:2012ja,Cantele:2009de,novelli_optical_2020}.

From a geometric point of view, a generic twist angle of tBLG does not guarantee an exact in-plane periodicity, \textit{i.e.},  a commensurate structure. At some particular angles, however, tBLG forms commensurate structures. The moir\'e pattern arising from the twist induces a lattice periodicity, which defines the moir\'e Brillouin zone (MBZ) in its reciprocal space. A continuum model has been developed within the MBZ and widely used to study the electronic band structure of tBLG \cite{bistritzer_moire_2011,lucignano_crucial_2019}. At these special angles, one can use tight-binding models or even first-principles calculations with periodic boundary conditions. 

\begin{figure}[t!]
\centering
\includegraphics[width=0.5\textwidth]{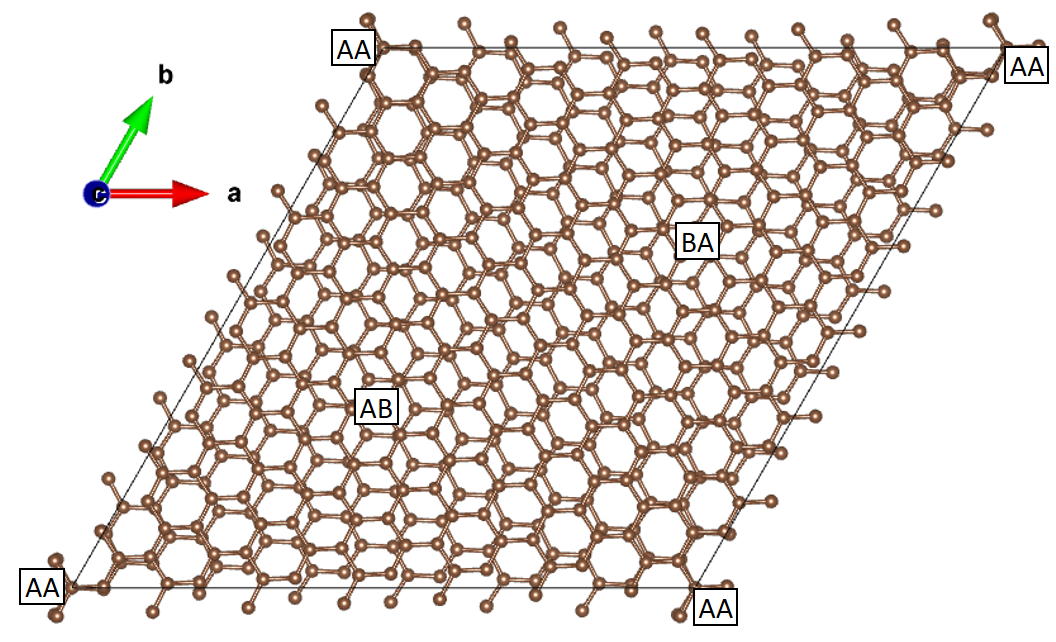}
\caption{
Supercell of tBLG with twist angle equal to $+5.086^\circ$ (R-chiral). AA- and AB/BA-stacked regions are labeled.
}
\label{fig:system1}
\end{figure}

The structural deformations in the mtBLG are one of the key factors in determining the gap between the flat bands manifold from the other bands~\cite{angeli_emergent_2018,choi_strong_2018,lin_shear_2018,lucignano_crucial_2019}. Structural relaxations can be calculated by using empirical potentials~\cite{wijk_relaxation_2015,jain_structure_2016,gargiulo_structural_2017,angeli_emergent_2018,choi_strong_2018,leconte_relaxation_2019} or by first-principles methods  ~\cite{uchida_atomic_2014,lin_shear_2018,lucignano_crucial_2019,cantele_structural_2020}. The tBLG at small twist angles shows different regions labeled as AA-stacked and AB/BA-stacked (Bernal-stacked) as shown in Fig.~\ref{fig:system1}. Each region is similar to the corresponding untwisted bilayer graphene. In the AA stacked bilayer graphene, all the sublattices of the top layer are superposed on those of the bottom layer, whereas in the AB-stacked bilayer, one sublattice of the top layer is on top of the carbon atom but the other sublattice is on top of the vacant hexagon center of the bottom layer. Several theoretical studies show vortices as in-plane displacements centered at the AA region  \cite{wijk_relaxation_2015,gargiulo_structural_2017,angeli_emergent_2018,choi_strong_2018,cantele_structural_2020}, where the in-plane displacements are related to atomic relaxations with respect to the ideal twisted system.
This vortex pattern tends to increase the area of the AB/BA region and to reduce the area of the AA region. 
On the other hand, as far as the out-of-plane direction is concerned, the AA region has a larger interlayer distance than AB/BA region \cite{uchida_atomic_2014,wijk_relaxation_2015,jain_structure_2016,choi_strong_2018,lucignano_crucial_2019}. All these observations are consistent with the result that AB/BA stacking is energetically favored with respect to AA stacking in the bilayer graphene.
Without structural relaxations, electronic structure calculations show that conduction and valence bands near the Fermi energy have a bandwidth, which appears larger than the experimental value~\cite{Kerelsky:2019aa,Xie:2019aa,Jiang:2019aa,Choi:2019aa}. Furthermore, they are not well separated from other bands as a necessary condition required by flat bands. By introducing structural relaxations in the mtBLG, the flatness and the separation from other bands become consistent with the experimental findings~\cite{choi_strong_2018,lucignano_crucial_2019}. This points out the essential role of atomic relaxations in electronic structure calculations for describing the physical properties of tBLG.

In this study, we discuss the interplay of chirality of the twist
with the structural relaxations which is discussed in real space as well as with the electronic and spin-orbit properties of the bilayer which are shown in momentum space. To the best of our knowledge, the chirality of the twist has been considered only in a
few works as far as the optical properties are concerned~\cite{kim_chiral_2016, suarez_morell_twisting_2017, stauber_chiral_2018, addison_twist_2019, stauber_change_2020,Do:PRR20}. 
Very recently, it turns out that the chirality can induce an asymmetry in the transport phenomena of the tBLG~\cite{liu_chirality-induced_2021}.
Also, the chiral supercurrent in the tBLG has been theoretically suggested~\cite{wu_topological_2019}.

When building the tBLG by twisting one layer of the original untwisted system, one has two possible choices, \textit{i.e.}, negative or positive twisting rotation with respect to a reference oriented \textit{z}-axis. In this work, we define the right (left) twist as the rotation of the top layer counter-clockwise (clockwise) where the bottom layer is fixed, and the rotation axis is the $z$-axis positively oriented from the bottom layer towards the top layer. We refer to the tBLGs obtained by the right and left twist as Right- (R-) chiral and Left- (L-) chiral tBLG, respectively. We also denote the Right (Left) twist by $\theta$ as $+\theta$ ($-\theta$) as a shorthand notation.

If the mirror operation is applied to the R-tBLG, where the mirror plane is the middle-plane between the top and bottom layer, one can switch from R- to L-chiral structure and viceversa~\cite{uchida_atomic_2014, kim_chiral_2016, stauber_chiral_2018}.  It implies that the L- and R-chiral tBLGs are twins or enantiomer pairs, thus justifying the use of the term `chirality'. To change the twist chirality in tBLG, one needs to decouple the two layers, twist them in the opposite way, and couple them again. It means that the two twist chiralities are structurally distinct and they are not connected by continuous structural deformations induced by external fields. Therefore, they should be physically distinguishable.
For instance, different chirality should induce different optical activity. Indeed,  L- and R-tBLG give rise to the opposite circular dichroism spectra, which is the difference in absorption between left- and right-handed circularly polarized light~\cite{kim_chiral_2016}.   
Two enantiomers tBLG should induce the opposite rotation of the polarization plane of linearly polarized light (natural optical activity), as it travels through the systems.
Moreover, this effect should depend on the twist angle, and on the atomic relaxation following the twist. This could open new directions to explore, such as optical activity engineering by a twist or by combining different twisted bilayers or multilayers with different chirality through van der Waals interactions, \textit{i.e.} heterostructure chirality engineering.  

Besides reproducing the reported in-plane vortex and out-of-plane corrugated patterns in the atomic displacements~\cite{uchida_atomic_2014,wijk_relaxation_2015,jain_structure_2016,gargiulo_structural_2017,angeli_emergent_2018,choi_strong_2018,lucignano_crucial_2019,cantele_structural_2020}, our calculations show that the helicity of in-plane displacement vortices is alternating in each layer when considering an L- or R-chiral tBLG. We use the term `helicity' to refer to the clockwise or counter-clockwise local rotation of the corresponding vector field, \textit{i.e.}, displacement field or spin texture represented by arrows. 
More interestingly, our calculations show an inversion of the helicity of the spin texture connected to a different chirality of tBLG,\textit{ i.e. }L-tBLG or R-tBLG.
The correlation between the spin texture and the structural chirality of the tBLG is a new interesting property, which is different from the switching of the spin texture helicity upon ferroelectric polarization reversal, occurring in bulk systems~\cite{di_sante_electric_2013, tao_reversible_2017, wei_discovery_2020}.
It is rather similar to the spin texture depending on the structural chirality in the tellurium crystal~\cite{sakano_radial_2020}.

Several theoretical studies of the spin texture in the untwisted bilayer graphene can be found in the literature~\cite{winkler_electromagnetic_2015, gmitra_proximity_2017}.
Recently, spin texture in the twisted magnetic Janus bilayer has also been reported~\cite{soriano_spin-orbit_2021}.
Similar to our study, Ref.~\cite{shen_exotic_2021} introduced the chirality of the twist to study the magnetic twisted bilayer system and chiral dependence of the ferroelectric property of it. However, their definition concerns the $30^\circ$ twist, which is the singularity where the moir\'e periodicity disappears.

\section{Description of the systems}
To describe the structure of the tBLG, we follow the method used in Ref.\ \cite{lucignano_crucial_2019,moon_optical_2013}. The primitive cell of single-layer graphene consists of the Bravais lattice vectors $\mathbf{a}_1 = a(\sqrt{3}/2,-1/2)$ and $\mathbf{a}_2 = a(\sqrt{3}/2,1/2)$ where $a=2.456$ \AA. One carbon atom is placed in the origin, and the other is at $(1/3)(\mathbf{a}_1+\mathbf{a}_2)$. For the twist angle $\theta>0$ corresponding to the right twist, the primitive lattice vectors of each layer are rotated as $\mathbf{a}_{i}^{(l)}=\mathsf{R}(\pm\theta/2)\mathbf{a}_{i}$, where $l=t,b$ refers to the top and bottom layer respectively. $\mathsf{R}(\theta)$ is the rotation operator by angle $\theta$. If $\theta$ satisfies the condition $2\cos\theta = (m^2+n^2+4mn)/(m^2+n^2+mn)$ for a pair of integers $(n,m)$, tBLG respects the periodicity, \textit{i.e.,} it has a  commensurate structure~\cite{shallcross_electronic_2010}. In this case, the sign of $\theta$ is plus (minus) if $n>m\ (n<m)$. The lattice vectors of the supercell defined by $(n,m)$ are $\mathbf{L}_1 = n\mathbf{a}_{1}^{(t)}+m\mathbf{a}_{2}^{(t)} = m\mathbf{a}_{1}^{(b)}+n\mathbf{a}_{2}^{(b)} = (L,0)$ and  $\mathbf{L}_2 = \mathsf{R}(\pi/3)\mathbf{L}_1$ where $L = a\sqrt{n^2+m^2+mn}$ is the supercell lattice constant.
The interlayer distance of the unrelaxed system is $d_0 = 3.348$ \AA. The lattice constant of out-of-plane direction for the supercell is chosen as $c = 10$ {\AA} to include the vacuum region and prevent the interaction between the periodic spurious replicas along the out-of-plane direction.
By construction, the system has 3-fold rotational symmetry around the $z$-axis and 2-fold rotational symmetry around the three axes in the plane, one of which is along the $\mathbf{L}_1$ lattice vector.
We can easily construct the L-tBLG by inverting the sign of $\theta$ and exchanging $n$ and $m$ in the expression of $\mathbf{L}_1$. This is equivalent to  mirroring the top and bottom layers of the R-tBLG with respect to the middle plane between the two layers. Reciprocal lattice vectors generating the supercell Brillouin zone (SBZ) are $\mathbf{b}_{1}^{s} = (2\pi/L)(1,-1/\sqrt{3})$ and $\mathbf{b}_{2}^{s} = (2\pi/L)(0,2/\sqrt{3})$. 
In some cases, the SBZ does not coincide with the MBZ. We always adopt the SBZ in this work. 
To distinguish the valley degrees of freedom of $K$ points in the SBZ, we labeled $\mathbf{K_1} = (2/3)\mathbf{b}_{1}^{s} + (1/3)\mathbf{b}_{2}^{s}$ and $\mathbf{K_2} = -\mathbf{K_1}$. 

\begin{table}[b]
\centering
\begin{tabular}{c|cccc}
\hline
$\theta$ & $(n,m)$ & $L$ ({\AA}) & number of atoms & $k$-point grid \\
\hline
\hline
 $5.086^\circ$ & $(7,6)$ & $27.678$ & $508$ & $3\times3\times1$\\
 $9.430^\circ$ & $(4,3)$ & $14.939$ & $148$ & $5\times5\times1$\\
 $13.17^\circ$ & $(3,2)$ & $10.705$ & $76$ & $9\times9\times1$\\
 $16.43^\circ$ & $(5,3)$ & $17.192$ & $196$ & $5\times5\times1$\\
 $21.79^\circ$ & $(2,1)$ & $6.498$ &  $28$ & $9\times9\times1$\\
 $26.01^\circ$ & $(7,3)$ & $21.829$ & $316$ & $3\times3\times1$\\
 $29.41^\circ$ & $(8,3)$ & $24.189$ & $388$ & $3\times3\times1$\\
\hline
\end{tabular}
\caption{\label{tab:table1}%
Supercell information for tBLG with different twist angles.}
\end{table}

The supercell of the first magic angle, $\theta \sim 1.08^\circ$, contains 11164 atoms, which is too large for investigating the spin texture by using Density Functional Theory (DFT). Hence we adopt much smaller supercells which correspond to larger twist angles. For a  detailed analysis, we mainly focus on the tBLG with $\theta=\pm 5.086^\circ$. These tBLGs are generated by $(n,m) = (7,6)$ and $(6,7)$ for R- and L-chiral respectively. The R-chiral one is shown in Fig.~\ref{fig:system1}. They have 508 carbon atoms and a supercell lattice constant $L=27.678$ \AA. Furthermore, we also explore other twist angles focusing on the change of spin textures with the twist angle. Details for the supercells of these systems are listed in Table~\ref{tab:table1}.

\section{Method}
The structure relaxations of tBLG were done by using DFT as implemented in the Vienna Ab-initio Simulation Package (VASP)~\cite{kresse_efficient_1996}. Computational parameters are the same as used in  Ref.~\cite{lucignano_crucial_2019} except for the $k$-point samplings of the SBZ. Projector augmented wave pseudo-potentials~\cite{kresse_ultrasoft_1999} are adopted to represent the atomic cores. The energy cut-off for the plane-wave basis has been set to 400 eV. The $k$-point samplings in the self-consistent calculation are chosen according to the grids listed in Table~\ref{tab:table1}, all including $\Gamma$ point. 
In order to account for the van der Waals interactions between the atoms, the rev-vdW-DF2 exchange-correlation functional \cite{hamada_van_2014} is adopted. Structural relaxations are performed until the maximum magnitude of the force is smaller than 0.002 eV/\AA.

To investigate the electronic band structure, spin texture, and layer weight, we performed non-self-consistent calculations. The charge densities are obtained from the self-consistent calculation with the symmetrized structure under the $10^{-9}$ eV convergence criterion.
To calculate the spin texture, we performed non-collinear DFT calculations including spin-orbit coupling (SOC), using the charge density obtained from the collinear calculation, which is justified by tiny SOC strength.
Spin texture is calculated by the formula $\mathbf{s}_{n\mathbf{k}} = \frac{1}{2} \langle \psi_{n\mathbf{k}}| {\boldsymbol{\sigma}}|\psi_{n\mathbf{k}}\rangle$ where $\psi_{n\mathbf{k}}$ is the spinor Bloch wavefunction of band $n$ and $\boldsymbol\sigma$ is the Pauli matrices vector $\boldsymbol\sigma=(\sigma_x,\sigma_y,\sigma_z)$. Atomic units in which $\hbar = 1$ are adopted.
The layer weight is calculated by integrating $|\psi_{n\mathbf{k}}|^2$ over the volume defined by the middle plane of the bilayer and the upper (or lower) part of the supercell.
To represent the wavefunctions in real space, WaveTrans code is used~\cite{feenstra_low-energy_2013}.

\begin{figure*}[t!]
\centering
\includegraphics[width=\textwidth]{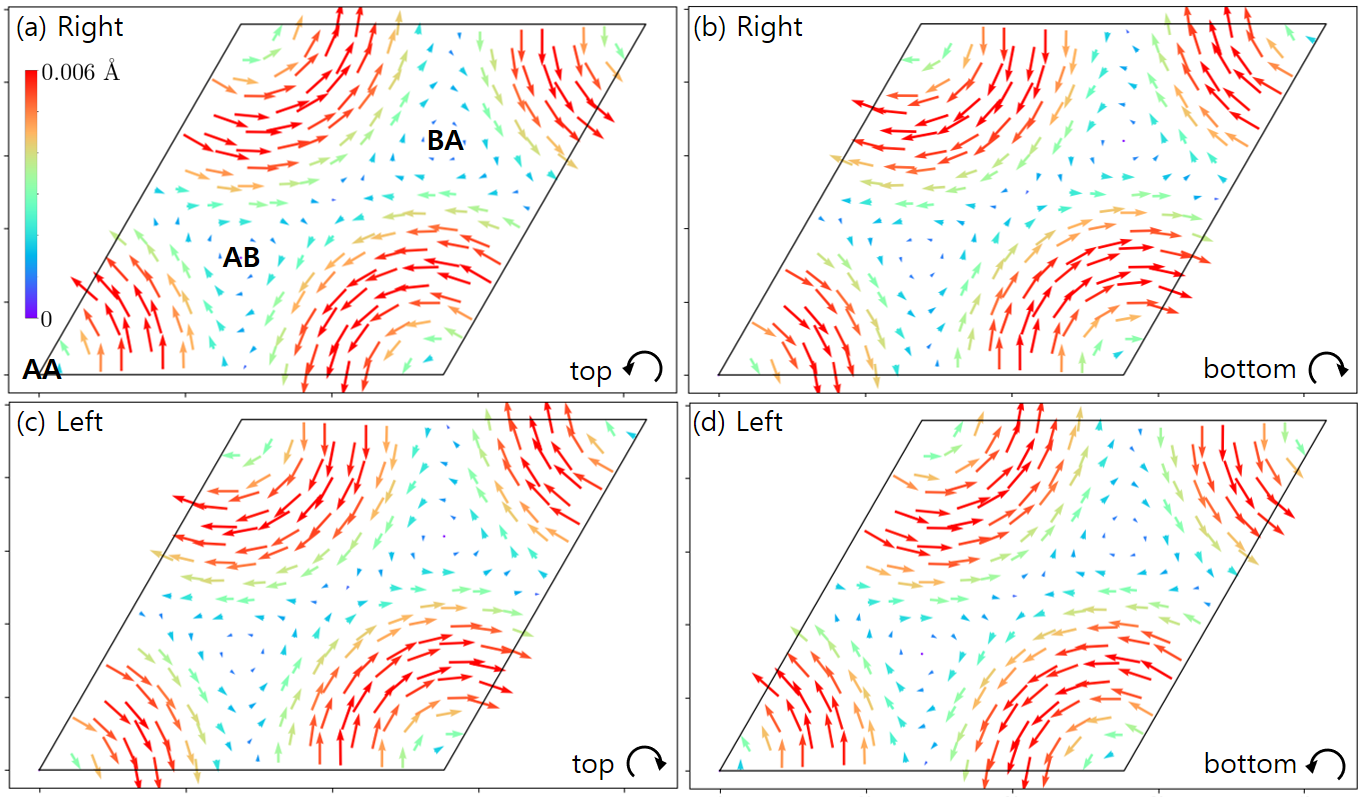}
\caption{In-plane displacements of atoms in the (a) top and (b) bottom layer of tBLG with $+5.086^\circ$ twist (R-chiral). The same for tBLG with $-5.086^\circ$ twist (L-chiral) are shown
in (c) and (d). Arrows in the cell show the direction of displacements of the atom placed at the tail of each arrow. The length of the arrows represents the relative size of the displacements. The color of the arrow represents the magnitude of displacement according to the colormap in (a). Round arrows at the lower right side of the panels represent the relative rotation of the layer with respect to the other layer.}
\label{fig:displ_inplane1}
\end{figure*}

\section{Results}
\subsection{Atomic displacements patterns}
As a first step, we consider the atomic relaxations in the tBLG with $\theta=\pm 5.086^\circ$. The displacement pattern is defined as the difference vectors of the relaxed atomic positions with respect to the unrelaxed twisted ones. In Fig.~\ref{fig:displ_inplane1} (a) we show the in-plane vortex-like displacement pattern of the top layer for the R-tBLG as discussed in Ref.~\cite{wijk_relaxation_2015,gargiulo_structural_2017,angeli_emergent_2018,choi_strong_2018,cantele_structural_2020}. The energy gain per atom by the relaxation is $\Delta E/N_{\text{atom}} \approx 0.48 ~\text{meV}$ corresponding to the thermal energy at $5.57~\text{K}$ which is consistent with Ref.~\cite{cantele_structural_2020}. Around the AA-stacked region, the displacement vector field shows a counter-clockwise vortex texture having the center of the region as a core. The intermediate region, including the AB/BA-stacked region, has minimal displacements compared with the AA-stacked region because they are close to an equilibrium configuration such as Bernal stacking. On the other hand, the in-plane displacement pattern of the bottom layer shows the clockwise vortex pattern around the AA-stacked region [see Fig.~\ref{fig:displ_inplane1} (b)]. This opposite helicity is also consistent with Ref.~\cite{jain_structure_2016,cantele_structural_2020}. In the L-tBLG, however, every vortex in each layer has the opposite helicity with respect to the corresponding one of the R-tBLG, \textit{i.e.}, the switching of the helicity of the in-plane displacement vortex occurs between the R- and L-tBLG.
Indeed, the top layer has a clockwise vortex-like pattern, and the bottom layer has a counter-clockwise vortex-like pattern, as shown in Fig.~\ref{fig:displ_inplane1} (c) and (d).

\begin{figure*}[!t]
\centering
\includegraphics[width=\textwidth]{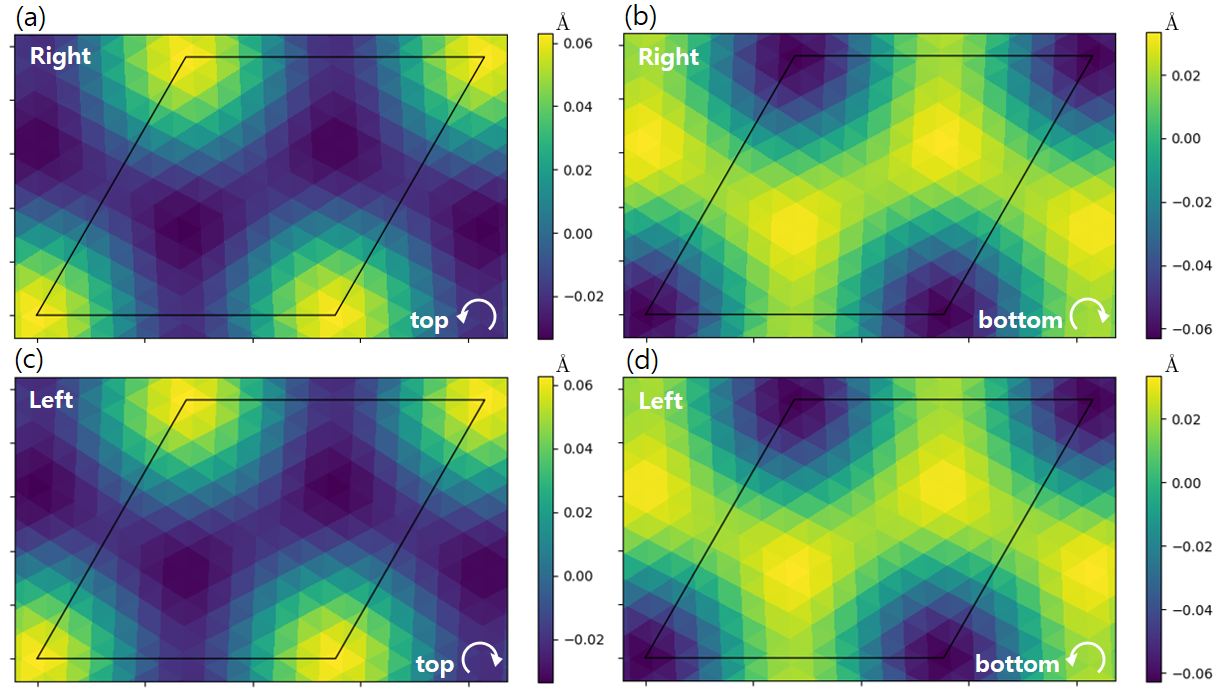}
\caption{Color map of the out-of-plane displacements of atoms in (a) top and (b) bottom layer of R-chiral tBLG with $+5.086^\circ$ twist. The unit of the displacements shown in the colorbars is \AA.
The same quantities are shown in (c) and (d) for L-chiral tBLG with $-5.086^\circ$ twist.
Round arrows at the lower right side of the panels represent the relative rotation of the layer with respect to the other layer.}
\label{fig:displ_z1}
\end{figure*}

Before introducing the twist, the ideal bilayer is in the energetically unfavored AA-stacked phase where the carbon atoms are on unstable equilibrium positions. After the twist, carbon atoms in the AA-stacked region are slightly misaligned from atoms in the other layer. Then the repulsion between the atoms increases the misalignment because the initial reference positions correspond to an unstable equilibrium configuration. This is why the helicity of the atomic displacements in the top and bottom layers are opposite. 
For the same reason, displacements occur in the same direction of the relative twist of each layer, \textit{i.e.}, (counter-) clockwise displacement for the (counter-) clockwise relative twist of each layer. This explains also why the helicities in the L- and R-tBLGs are opposite.

We also investigate the out-of-plane displacements, \textit{i.e.}, along the $z$-axis. The displacements are measured from the average value of the $z$-components of atoms in each layer. At the top layer of the R-tBLG [Fig.~\ref{fig:displ_z1} (a)], the AA-stacked region marked in yellow-green color is the `hill' corresponding to the highest region in the layer. The AB/BA-stacked region marked in dark blue is the `valley', which is the lowest region in the layer. In the bottom layer, on the other hand, the AA-stacked region is the valley, and AB/BA-stacked region is the hill [Fig.~\ref{fig:displ_z1} (b)]. The interlayer distance is larger in the AA-stacked region due to the stronger repulsion between two layers as reported in Ref.~\cite{uchida_atomic_2014,wijk_relaxation_2015,jain_structure_2016,choi_strong_2018,lucignano_crucial_2019, cantele_structural_2020}. The out-of-plane displacement pattern is similar for both the L- and R-tBLG, as shown in Fig.~\ref{fig:displ_z1} (c) and (d). The chirality change does not affect the displacement pattern in the out-of-plane direction because it is mainly related to the AA- or AB/BA- stacking in the different regions. 

It should be noted that the concept of AA or AB/BA stacked region is well defined only at small twist angles or equivalently at angles close to $60^\circ$ or a multiple of it. Therefore, the in-plane vortex and out-of-plane hill and valley displacement patterns can only be defined in this condition [see Supplemental material (SM) Fig. S1~\cite{Suppl}].

\begin{figure}[t!]
\centering
\includegraphics[width=0.5\textwidth]{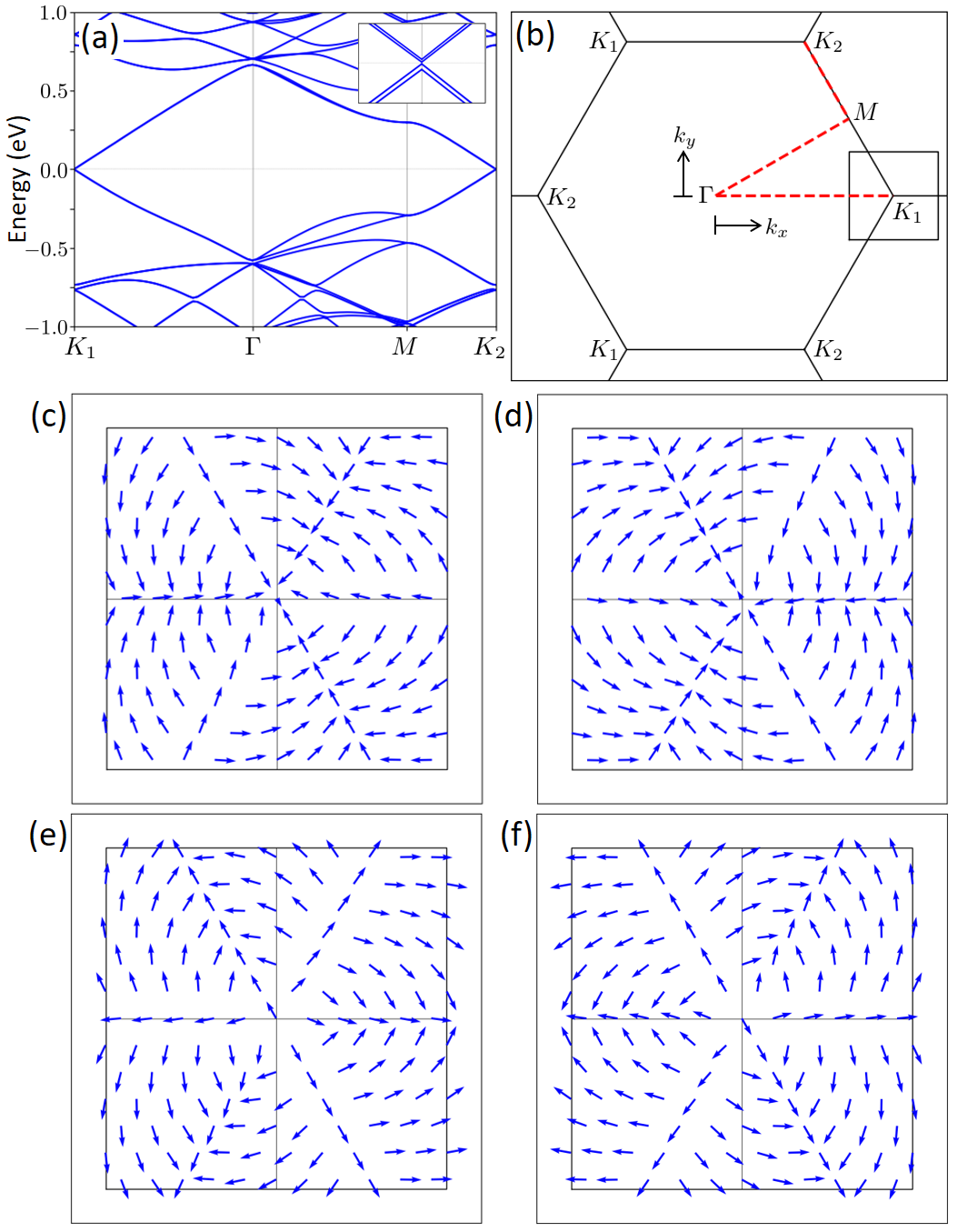}
\caption{
(a) Band structure and (b) SBZ with high symmetry points and band path of $5.086^\circ$ tBLG. Inset in (a) is a  zoomed portion of bands around $K_1$ point of $\Gamma$-$K_1$-$M$ path in the [-0.1,0.1] meV range. Zero energy corresponds to the Fermi level. The spin texture is calculated in the small rectangular region around the $K_1$ point in (b). 
Spin textures of R-tBLG at $+5.086^\circ$ at (c) $K_1$ and (d) $K_2$ points are shown. The same quantities of L-tBLG at $-5.086^\circ$ are shown in (e) and (f). If one inverts the spin in one figure, the spin texture in the other row and the same column is obtained. If one rotates the spin texture by 180$^\circ$ around an axis perpendicular to the figure, passing through K1 (or K2), it becomes the one in the other column and the same row.}
\label{fig:spintexture1}
\end{figure}

\subsection{Spin texture in momentum space}

Pristine single-layer graphene has Dirac points, which are the crossing points of linear band dispersion at $K$ points in the reciprocal space at the Fermi energy. The band structure of tBLG with $5.086^\circ$ twist angle is similar to that of the single-layer graphene, as shown in  Fig. \ref{fig:spintexture1} (a).
We now focus on the spin texture of the relaxed tBLG around the $K$ points. Although the tBLG is a non-magnetic system and the sizes of SOC band splittings are smaller than 1.5 $\mu$eV at $K$ points in our calculations, the spin textures are well defined at generic $\mathbf{k}$-point, since they represent the mean-value of spin operators. Two spin split bands due to the SOC have exactly opposite spin directions at the same $\mathbf{k}$ [SM Fig. S3].
In Fig.~\ref{fig:spintexture1} (c), the spin texture of the highest valence band around the $K_1$ point of the R-tBLG is shown. Spin textures are calculated in the rectangular region shown in Fig.~\ref{fig:spintexture1} (b) where the length of the side is 0.075 \AA$^{-1}$.  

The spin texture is divided into six regions according to which direction the spins are pointing at. Each region can be seen as a part of a vortex, although slightly canted inward. The helicity in one region is opposite to the neighboring one. On the boundaries between these regions, spins are pointing to the $K_1$ point.
The topology of these spin-textures is completely new to the best of our knowledge. In particular, they belong neither to the Rashba nor the Dresselhaus topology \cite{0034-4885-78-10-106001}. So it deserves more attention, as we will discuss shortly. The magnitude of each spin vector is almost $1/2$, and the out-of-plane component is nearly 0. $K_1$ point is an exception having a much smaller spin.
Similarly to the $K_1$ point, around the $K_2$ point as shown in Fig.~\ref{fig:spintexture1} (d), the spin texture is divided into six regions with alternating helicity. Interestingly, the spin texture around $K_2$ is equivalent to a $180^\circ$ rotated image of the spin texture around $K_1$.
For the L-tBLG, spin textures around $K_1$ and $K_2$ [Fig.~\ref{fig:spintexture1} (e) and (f)] also show the same partition in six regions. However, spins are canted outward and the helicity is opposite. Spins on the boundary are outward direction. These cases correspond precisely to an inversion of the spins in the R-chiral cases. The relation between the spin textures around $K_1$ and $K_2$ points are the same as before, \textit{i.e.} $180^\circ$ rotated image. 

\begin{figure}[t!]
\centering
\includegraphics[width=0.5\textwidth]{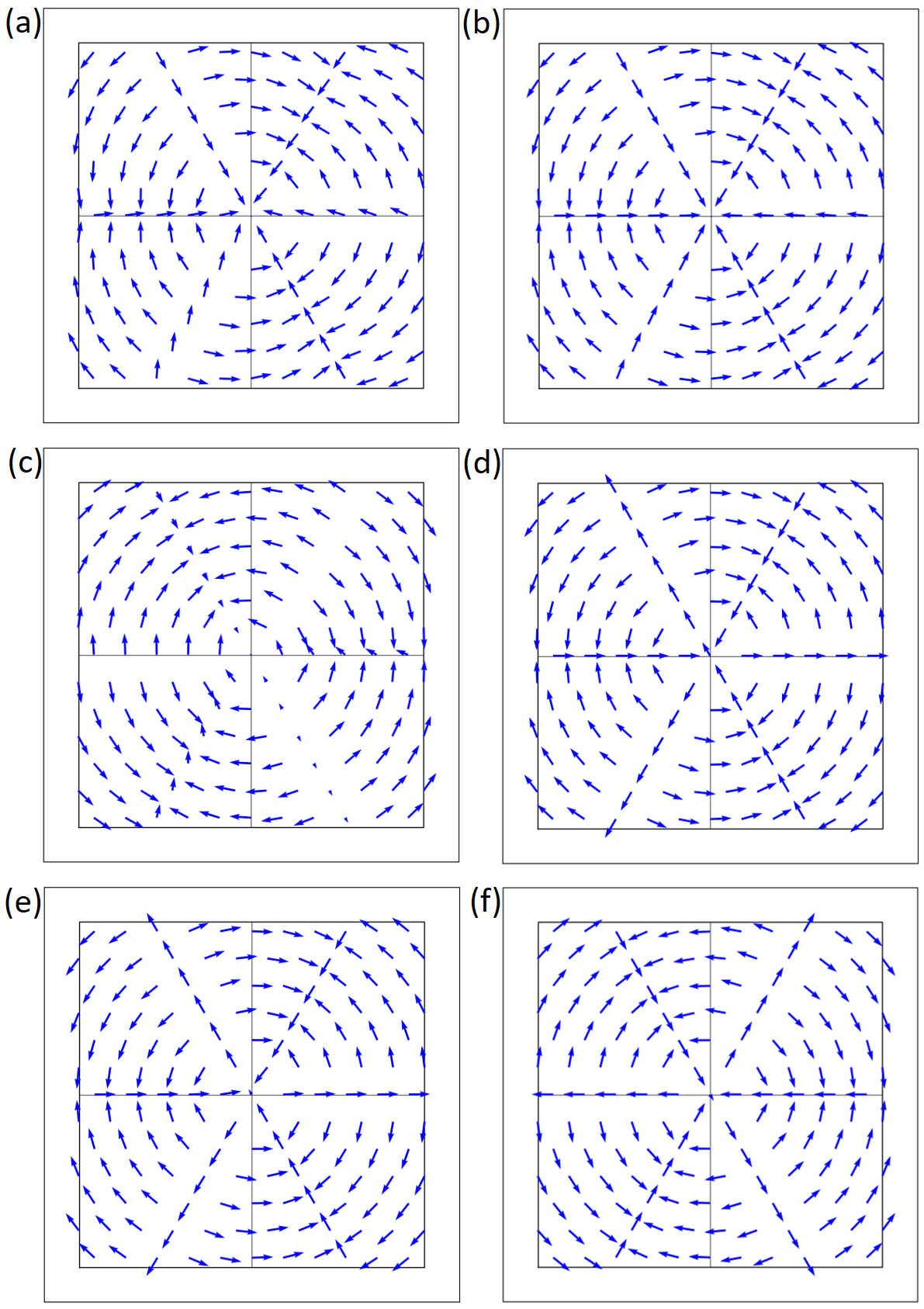}
\caption{Spin textures of tBLG with several twist angles at $K_1$: (a) $+9.430^\circ$, (b) $+13.17^\circ$, (c) $+16.43^\circ$, (d) $+21.79^\circ$, (e) $+26.01^\circ$, and  (f) $+29.41^\circ$.}
\label{fig:spintexture3}
\end{figure}

The spin texture of tBLG should obey the $D_3$ point group symmetry with respect to the $K$ points, \textit{i.e.}, 3-fold rotation around the out-of-plane axis and 2-fold rotations around in-plane axes. The boundaries between the spin texture regions coincide with the 2-fold symmetry axes. On a 2-fold axis, the only allowed spin direction is parallel to the axis by the symmetry~\cite{cheong_sos_2019}. 
However, some spins on the 2-fold axes slightly show some deviations due to the numerical uncertainty. We will come back to this point later.

We also investigate the spin textures for different twist angles corresponding to systems with different sizes.  In Fig.~\ref{fig:spintexture3}, spin textures of the highest valence band around $K_1$ point of the R-tBLGs with the twist angles listed on Table~\ref{tab:table1} are shown. 
For comparison, spin textures are calculated in the $k$-space region of the same area as that used for the $5.086^\circ$ case of Fig.~\ref{fig:spintexture1}. We confirmed that the inversion of spins by the change of twist chirality and the $180^\circ$ rotation between the spin vector fields at $K_1$ and $K_2$ points are general properties of tBLG [See SM Fig. S6]. For larger twist angles, spin texture exhibits vortex-like regions with an alternating helicity, as shown in Fig.~\ref{fig:spintexture3}, similarly to the $5.086^\circ$ case. Spins are also canted from the perfect circular direction. The deviations are more apparent for smaller twist angles.
Interestingly, spins on the 2-fold axes (radial spins) change their behavior as the 
twist angle changes. In Fig.~\ref{fig:spintexture3} (a), the spins on the 2-fold axes show some deviations 
from the axis direction, in some cases. This is more evident in (c), which looks even more special, 
since in some radial directions,  the spins either disappear or become very small in magnitude. In (b), the orientations of the radial spins are all inward with respect to $K_1$, but collinear with the symmetry axis. In all the other cases, (d-f), we have three outward and inward radial spins.

Interestingly, the spin textures with large twist angles have a topology of winding number $+4$, that is, if one draws a closed path including the $K_1$ point and moves along it counter-clockwise, spin on this path rotates counter-clockwise four times. On the other hand, spin textures with smaller twist angles have a winding number $+1$.
This suggests that there is a transition in the topological properties of the spin texture from large to small twist angles, where 
the $16.43^\circ$ twist angle could represent a case close to the `transition angle' where the spin texture is extremely sensitive to small numerical noise. Unfortunately, our computational resources do not allow us to benchmark this interpretation.
We remark that even the increased computational parameters such as $k$-point grid or plane-wave energy cut-off could not improve the spin texture at $16.43^\circ$.

The spin textures of the conduction bands are reported in the SM, see Fig. S4. They are similar to those of valence bands and have the same switching rule.
Because the chirality is induced by the twist, spin texture arises even without the relaxation.  We can observe that the relaxation induces small changes in the spin texture [SM Fig. S8].

\begin{figure}[t!]
\centering
\includegraphics[width=0.5\textwidth]{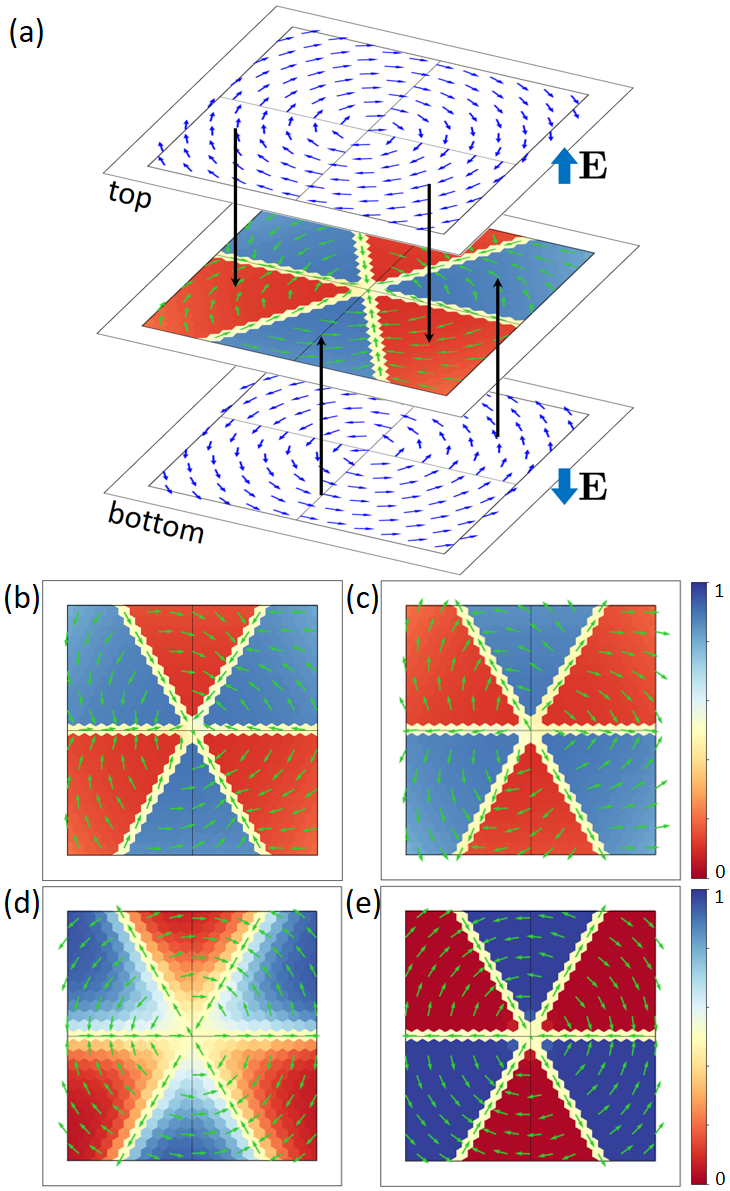}
\caption{(a) Schematic plot showing the origin of the alternating spin textures. The spin texture of the single-layer graphene in the presence of the (upper panel) upward and (lower panel) downward out-of-plane ($z$) external electric field corresponding to the top and bottom layers, respectively. The middle panel of (a) is the same as (b). In this case, the top and bottom layer dominant regions are indicated by the black arrows from the corresponding single-layer spin textures. (b-e) The bottom layer weight distribution of the highest valence band of the (b) $+5.086^\circ$, (c) $-5.086^\circ$, (d) $+21.79^\circ$, and (e) $+29.41^\circ$ tBLG around the $K_1$ point are also shown. The same region as in Fig.\ \ref{fig:spintexture3} is shown.}
\label{fig:Rashba1}
\end{figure}

The projection on each layer of a Bloch wavefunction at each $\mathbf{k}$, which is the probability for the Bloch electron to be localized in a specific layer, can be useful to study the origin of these novel spin textures of the tBLG. For the R-chiral $+5.086^\circ$ tBLG, the weight of the bottom layer of the highest valence band around the $K_1$ point is shown by the color map in Fig.~\ref{fig:Rashba1} (b). Blue (red) means that the bottom (top) layer contributes with a larger weight. Remarkably, the top dominant and bottom dominant regions are clearly separated. The layer contributing with the larger weight changes suddenly at the 2-fold axes which define the boundaries of the regions. One would expect that the layer weight distribution is also related to the $D_3$ symmetry. Indeed, the 3-fold rotation symmetry is clearly seen. The 2-fold rotation exchanges the top and bottom layer so that the layer weight is also inverted across the 2-fold axes, which represent equally weighted lines.
One can see that the partition of the spin texture discussed previously coincides with the partition according to the dominant layer-weights. This suggests that the spin texture at a specific $k$-point has a ``character'' mainly associated with either top or bottom layer.

The origin of the spin texture must be traced back to the spin-orbit interaction. Rashba effect refers to the SOC  splitting of the spin degenerated bands induced from the inversion symmetry breaking of the potential due to the crystal symmetry, to the external electric field, or to the substrate~\cite{Rashba_1960,zhang_hidden_2014,0034-4885-78-10-106001,kane_z_2_2005}. A circular spin texture is often found in these cases. Rashba effect in graphene has been studied using the tight-binding theory in Ref.~\cite{kane_z_2_2005,min_intrinsic_2006}. 

In tBLG, one can fictitiously consider the following decomposition: 
the top layer can be seen as single-layer graphene above the bottom layer which plays the role of the substrate, and vice versa. 
Therefore, each layer is subject to a  Rashba effect arising from the potential gradient generated by the other layer. 

At this point, one can investigate the spin texture of the isolated single-layer graphene by the Rashba effect induced by an external electric field instead of considering the potential gradient generated by the other layer. 
In our case, it is possible to simulate the presence of an external electric field, implemented as a sawtooth type potential in DFT~\cite{neugebauer_adsorbate-substrate_1992}.
In Fig.~\ref{fig:Rashba1} (a) upper and lower panels show the calculated spin textures of the highest valence band of the single-layer graphene in the presence of an external electric field, oriented along the $z$ direction, of $0.01$ eV/{\AA} and $-0.01$ eV/{\AA}, respectively. These values are chosen arbitrarily. These circular spin textures are characteristic of Rashba-type textures. The upward electric field induces a clockwise Rashba spin texture which becomes counter-clockwise when the external electric field is switched and vice versa. 
After these preliminary considerations for the single-layer graphene under a fictitious external electric field, we now turn our attention to the twisted bilayer graphene.

In  Fig.~\ref{fig:Rashba1} (b), we consider the spin texture and the layer weight distribution of the tBLG. The blue and red regions correspond to the dominant weight on the bottom and top layers respectively. Interestingly, the blue and red regions show a spin texture with dominant counter-clockwise and clockwise circular texture, respectively. 
The spin texture of the single-layer graphene under the external field [Fig.~\ref{fig:Rashba1} (a)]  suggests that the bottom (top) layer corresponds to the layer to which the downward (upward) electric field is applied. This correspondence is represented schematically in Fig.~\ref{fig:Rashba1} (a). 
One can see the same property in other cases in Fig.~\ref{fig:Rashba1} (c-e) apart from small deviations from the perfect circular texture.
Therefore, one can conclude that the spin textures of tBLG can be decomposed in different contributions of the Rashba spin texture of the top or the bottom layers.

However, small deviations of the spins with respect to the radial directions, \textit{i.e.} 2-fold axis directions, could be due to numerical noise since the change in the layer weight is quite drastic near the 2-fold axes.

Interestingly, the weight distribution of the top and bottom layers also obeys a switching rule similar to the spin texture. Fig.~\ref{fig:Rashba1} (c) is the layer weight of the L-chiral $-5.086^\circ$ tBLG. The top and bottom dominant regions are exchanged in comparison with the R-tBLG so that the corresponding Rashba spin texture helicity is also exchanged. It is in agreement with the enantiomer relation between the two systems, \textit{i.e.}, the L-tBLG is obtained by exchanging the top and bottom layer of the R-tBLG, and the top and bottom dominant regions are also exchanged.
The layer weight distribution around the $K_2$ point is a $180^\circ$ rotated image of that around the $K_1$ point [SM Fig. S5]. The switching of the layer weight distribution supports our explanation for the spin textures.

The layer weight distribution of the large-angle case of $+29.41^\circ$ is shown in Fig.~\ref{fig:Rashba1} (e). The features are the same as in the small-angle case of $+5.086^\circ$ except for the inverted top and bottom dominant regions [see SM Sec. S\textrm{V} and Fig. S12]. Furthermore, the darker colormap suggests that the top and bottom layers are less mixed in this case.
An interesting case is $+21.79^\circ$ shown in Fig.~\ref{fig:Rashba1} (d). Unlike the other twist angles for which the top and bottom dominant regions are clearly separated, the boundary region in Fig.~\ref{fig:Rashba1} (d) changes quite smoothly. The reason why only the $+21.79^\circ$ case exhibits this feature is not clear.

\begin{figure}[t!]
\centering
\includegraphics[width=0.5\textwidth]{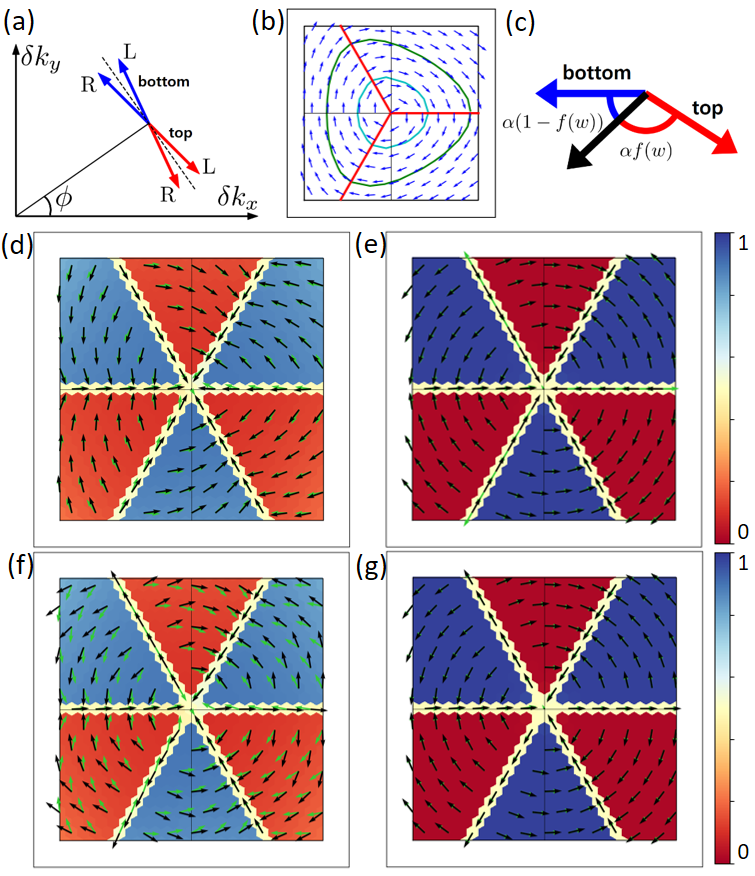}
\caption{(a) The schematic expression of the $F^{\text{(t/b)}}_{1,\theta}$ of the model 1. The red and blue arrows mean the spin of the top and bottom layers, respectively. (b) Trigonal warping of the spin texture in the single-layer graphene around $K_1$ in the region of 0.45 \AA$^{-1}$ side length. The red solid line means the Brillouin zone boundary. The bright blue and green curves are equivalent energy lines of -0.5 and -1.0 eV. (c) The mixing rule by $f(w)$ determining the spin direction from the spin of the top and bottom layers and the weight $w$. (d-g) Model spin texture (black arrows), the calculated spin texture (green arrows), and the bottom layer weight distribution (color map) of (d) Model 1 $+5.086^\circ$, (e) Model 1 $+26.01^\circ$, (f) Model 2 $+5.086^\circ$, and (g) Model 2 $+26.01^\circ$.}
\label{fig:model_plot1}
\end{figure}

In order to further verify our interpretation, we construct two plot regression models reproducing the spin texture of tBLG. They consist of the model spin texture for each top and bottom layer and the mixing rule for the spin with respect to the bottom layer weight $w$. At a given $\textbf{k}$, the top layer spin $\mathbf{s}_{t}$ and the bottom layer spin $\mathbf{s}_{b}$ are determined by the single-layer model discussed above. Then, the resulting spin of tBLG is determined from $\mathbf{s}_{t}$ and $\mathbf{s}_{b}$ by a suitable mixing rule.
At first sight, one may intuitively think that the resulting spin is a linear combination of two spins with the layer weight as coefficients, $(1-w)\mathbf{s}_{t}+w\mathbf{s}_{b}$. However, we need to modify
this intuitive interpretation as follows. In our spin texture, the magnitudes of the spins are almost $1/2$ everywhere. However, the linear combination of the spins will have a much smaller magnitude than $1/2$ in general. This is why the mixing rule for the spin can not be represented as a linear combination.
Instead, as a mixing rule, we follow an isometry, \textit{i.e.} a transformation which 
preserves the module of the vectors. Specifically, we use a rotation.
At a given $\textbf{k}$, $\mathbf{s}_{t}$ and $\mathbf{s}_{b}$ define an angle $\alpha$ which satisfies $\cos\alpha=\mathbf{s}_{t}\cdot\mathbf{s}_{b}/(|\mathbf{s}_{t}||\mathbf{s}_{b}|)$ and $-180^\circ \leq \alpha \leq 180^\circ$. $\alpha$ is positive when $\mathbf{s}_{t}$ is rotated counter-clockwise with respect to $\mathbf{s}_{b}$. 
We suggest that the resulting spin is $\mathsf{R}((1-f(w))\alpha )\mathbf{s}_{b}$, 
pointing along the direction obtained from $\mathbf{s}_{b}$ by a rotation $\mathsf{R}$ by $(1- f(w))\alpha$, where $0\leq f(w)\leq 1$ is the mixing rule given by a function monotonously increasing as  $w$ increases. It is shown in Fig.~\ref{fig:model_plot1} (c). 

Let us first consider the case that the model spin texture of each layer is an ideally circular Rashba spin texture. Then the spins are collinear but pointing to  opposite direction  at every $\textbf{k}$, \textit{i.e.,} $\alpha=\pm 180^\circ$. The sign of $\alpha$ is not uniquely determined. 
To resolve the ambiguity, one can introduce a small deviation from the collinearity of the spins. In this way, $\alpha$ is no longer $\pm 180^\circ$ in the generic case. 

The model for the spin texture of each layer is described as $\phi_s^{\text{(t/b)}} = \phi \mp\pi/2 + F^{\text{(t/b)}}_{\theta}(\phi)$, where $\mp$ is for top or bottom layer. Here $\phi_s$ is the angle of spin direction, $\theta$ is the twist angle, and $\phi$ is the angle of the $\delta\textbf{k}$ which is crystal momentum measured from $K_i$ point. $\phi_s$ and $\phi$ are measured with respect to  the $k_x$-axis.  $F^{\text{(t/b)}}_{\theta}(\phi)$ is the spin deviation angle from the ideal Rashba spin texture in the $\theta$-twisted tBLG. 
Note that these models are not derived from the physical Hamiltonian but we propose as a non-linear plot regression model to reproduce the spin textures. We will show that the proposed models are in good agreement with the calculated spin texture.

Model 1 supposes that the spins are canted inward (outward) for the R-tBLG (L-tBLG) with respect to the tangential direction to the circle centered at $K$, see dashed line as shown schematically in Fig.~\ref{fig:model_plot1} (a).
As discussed before, the canting is larger for  the smaller twist angles, then  the following function is suggested
\begin{equation}
\begin{split}
F^{\text{(t/b)}}_{1,\theta} = \mp\text{sgn}(\theta)\frac{\pi}{2}\exp(-A_1\abs{\theta}) ,\\
\end{split}
\end{equation}
where sgn($\theta$) is $+1 (-1)$ for R(L)-tBLG and $A_1 = 16.015$ rad$^{-1}$ is a fitting parameter extracted from $+5.086^\circ$ tBLG.
The mixing rule function is assumed as $f_{1}(w)=(1/2)(1-\cos\pi w)$.
The spin texture of the $+5.086^\circ$ tBLG derived by model 1 is shown by black arrows in Fig.~\ref{fig:model_plot1} (d), while the green arrows represent the calculated spin texture.
The black and green arrows are almost overlapping suggesting that model 1 is able to reproduce the details of the spin texture. For example, the inward spin canting is well reproduced [see Fig.~\ref{fig:model_plot1} (d)], as well as the outward spin canting [see SM Fig. S9].
However, model 1 fails to reproduce the ``3-in 3-out" spin texture on the 2-fold axes in the case of large twist angles, \textit{e.g.}, $+26.01^\circ$,  as shown in Fig.~\ref{fig:model_plot1} (e) despite the fact the remaining regions are well fitted.

In order to reproduce the ``3-in 3-out" spin texture on the 2-fold axes, we introduce model 2 which 
is based on the trigonal warping of the spin texture reported in the bilayer~\cite{gmitra_proximity_2017} and the single-layer~\cite{rakyta_trigonal_2010,ma_trigonal_2018} graphene.
Fig.~\ref{fig:model_plot1} (b) shows again the case of single-layer graphene in the upward external field but in the larger area where the trigonal warping is clearly seen. The spin texture around the $K$ point is deformed to a triangle-like shape. We assume the trigonal warping with respect to the unfolded Brillouin zone of the top and bottom layers. In this case, the spin texture reproducing the trigonal warping can be given by
\begin{equation}
\begin{split}
F^{\text{(t/b)}}_{2,\theta}(\phi) &= A_2\sin(3(\phi\mp\text{sgn}(\theta)\pi/2\mp\theta/2)) ,\\
\end{split}
\end{equation}  
where the amplitude of the deviation $A_2=\pi/60$ is an arbitrary small parameter.
These textures should be rotated by $\pi$ for some twist angles which have different Brillouin zone folding patterns [SM Sec. S\textrm{V}]. The mixing function is assumed to be the same as that of model 1, $f_{2}(w)=(1/2)(1-\cos\pi w)$.
Fig.~\ref{fig:model_plot1} (g) shows that the spin texture derived from model 2 well reproduces the case of the $+26.01^\circ$ tBLG not only in the top or bottom dominant regions but also on the 2-fold axes. It implies that trigonal warping is necessary for reproducing the spin texture of the large angle tBLGs.
On the other hand, model 2 poorly reproduces the spin texture of $+5.086^\circ$ tBLG as shown in Fig.~\ref{fig:model_plot1} (f). It fails to reproduce the inward canting.

In summary, none of these two models reproduces in all the details the spin texture for all the twist angles. One can suppose that model 1 (2) is proper for the small (large) twist angles. The non-linear regression plot models for other cases are shown in SM Fig. S9 and S10.
Finally, we remark that in untwisted bilayer graphene, weights of both layers are equally distributed over the respective regions and spin texture is circular (AB) or does not appear (AA). It implies that our model can not be applied to the untwisted case.

\section{Discussion}
In general, a commensurate tBLG has  translational symmetry as imposed by the supercell lattice vectors. An arbitrary in-plane translation of one layer with respect to the other does not spoil the commensurability and it does not affect the continuum Hamiltonian at a small twist angle~\cite{bistritzer_moire_2011}. In some special cases, additional symmetry operations can increase the symmetry. For example, if the untwisted bilayer graphene is AA stacked bilayer and one twists it around the atomic site, the tBLG has $D_3$ point group symmetry corresponding to $P321$ space group (no. 150). This is the same starting configuration that we adopted in this study. If one twists it around the central point of a hexagon of the graphene, it has $D_6$ point group symmetry corresponding to the $P622$ space group (no. 177). The pattern of displacements shows similar characteristics in this case [SM Fig. S2].
Recently, emergent $D_6$ symmetry in the Bloch wavefunctions of low energy bands of mtBLG has been reported~\cite{angeli_emergent_2018} even if the system does not have this symmetry. 
A detailed analysis of the possible symmetries in the commensurate tBLG has been done in Ref.~\cite{zou_band_2018}. They defined the ``conjugated'' structure, which corresponds to our concept of the L-tBLG with respect to the R-tBLG, \textit{i.e.} enantiomers.
Interestingly, $D_3$ symmetry is preserved after the structural relaxations for all the tBLGs we have examined. For the magic angle 1.08$^\circ$ and some other small angles, $D_3$ symmetry breaking in the DFT calculation has been reported in Ref.\ \cite{lucignano_crucial_2019,cantele_structural_2020}.

The L-tBLG and R-tBLG form an enantiomer pair. By applying mirror operation with respect to the bilayer mid-plane, one can derive the L-chiral structure from the R-chiral structure, and vice versa,  because two layers are twisted in opposite directions for the two enantiomers. 
Because the $D_3$ point group does not have improper rotations such as mirror or inversion, L-tBLG and R-tBLG are not equivalent. However, the Hamiltonian of these structures  can be expressed as,
\begin{equation}
H_{R} = M_{z}^{\dagger}H_{L}M_{z}
\end{equation}
by the general transformation rule of the quantum mechanical operators.
$H_R$ and $H_L$ refer to the Hamiltonian of the R-tBLG and L-tBLG respectively. $M_z$ is the mirror operator which has a mirror plane parallel to the layer. 
From this relation and the eigenvalue equation
\begin{equation}
\begin{split}
H_{R}\psi^{R}_{n\mathbf{k}} &= E_{n\mathbf{k}}\psi^{R}_{n\mathbf{k}} \\
H_{L}(M_{z}\psi^{R}_{n\mathbf{k}}) &= E_{n\mathbf{k}}(M_{z}\psi^{R}_{n\mathbf{k}}), 
\end{split}
\end{equation}
we can relate the eigenstates of the R-tBLG $\psi^{R}_{n\mathbf{k}}$ and L-tBLG $\psi^{L}_{n\mathbf{k}}$,
\begin{equation}
\label{eq:psi_left}
\psi^{L}_{n\mathbf{k}'} = M_{z}\psi^{R}_{n\mathbf{k}},
\end{equation}
where $n$ is the band index and $\mathbf{k}$ is crystal momentum. Because the tBLG is 2D material, crystal momentum is defined in 2D $(k_x,k_y)$ and $M_z$ does not change $\mathbf{k}$, \textit{i.e.,} $\mathbf{k}'=M_{z}\mathbf{k}=\mathbf{k}$. 
As a result, spin expectation values are
\begin{equation}
\expval{\mathbf{s}}^{R}_{n\mathbf{k}} = 
\tfrac{1}{2}\mel{\psi^{R}_{n\mathbf{k}}}{\boldsymbol{\sigma}}{\psi^{R}_{n\mathbf{k}}} = 
\tfrac{1}{2}\mel{\psi^{L}_{n\mathbf{k}}}{M_{z}^{\dagger}\boldsymbol{\sigma}M_{z}}{\psi^{L}_{n\mathbf{k}}}.
\end{equation} 
Mirror operator $M_z$ can be decomposed into inversion $I$ and 2-fold rotation $C_{2z}$, $M_{z} = IC_{2z}$. In the spin space, $M_{z}=C_{2z}$ because the angular momentum is invariant under inversion.
The  relation between the spin components at certain $\mathbf{k}$ of the L-tBLG and R-tBLG are 
\begin{equation}
\label{eq:spin_enantiomer1}
\begin{split}
&\expval{s_{x(y)}}^{L}_{n\mathbf{k}} = -\expval{s_{x(y)}}^{R}_{n\mathbf{k}} \\
&\expval{s_{z}}^{L}_{n\mathbf{k}} = \expval{s_{z}}^{R}_{n\mathbf{k}}. \\
\end{split}
\end{equation}
When one considers the different chirality, spins are rotated by $180^\circ$ around the $z$-axis, \textit{i.e.}, in-plane spin components are inverted but the $z$-direction component does not change.

Crystal momentum close to the $\mathbf{K_1}$ point can be written as $\mathbf{K_1} +\delta\mathbf{k}$. Since time-reversal is a symmetry operation, we can exploit it here as $T(\mathbf{K_1} +\delta\mathbf{k}) = \mathbf{K_2} -\delta\mathbf{k}$ because $K_1$ and $K_2$ are a time reversal pair. Therefore, $T\psi_{n,\mathbf{K_1} + \delta \mathbf{k}} = \psi_{n,\mathbf{K_2} - \delta \mathbf{k}}$ if we assume a non-degenerate case.
Also, $T$ can be written as $T=i\sigma_{y}\mathcal{K}$ in the spin space where the $\mathcal{K}$ is the complex conjugation operator.
\begin{equation}
\label{eq:spin_timereversal1}
\begin{split}
\expval{\mathbf{s}}_{n,\mathbf{K_1}+\delta\mathbf{k}} &= 
\tfrac{1}{2}\mel{\psi_{n,\mathbf{K_1}+\delta\mathbf{ k}}}{\boldsymbol{\sigma}}{\psi_{n,\mathbf{K_1}+\delta\mathbf{k}}} \\
&= \tfrac{1}{2}\mel{\psi_{n,\mathbf{K_2}-\delta\mathbf{ k}}}{T^{-1}\boldsymbol{\sigma}T}{\psi_{n,\mathbf{K_2}-\delta\mathbf{k}}}\\
&=-\expval{\mathbf{s}}_{n,\mathbf{K_2}-\delta\mathbf{k}}
\end{split}
\end{equation}
From this relation, if we rotate both the relative momentum $\delta\mathbf{k}$ and the spin direction around $K_1$ by $180^\circ$, we can obtain the spin texture around $K_2$.
In these derivations, the gauge freedom is neglected in Eq.~(\ref{eq:psi_left}) and the similar equation in the time-reversal part because it will be canceled when we calculate the spin texture. 

Orbital angular momentum operators, in general, can replace the spin operators in our discussion. Therefore, we expect that the switching rules of spin texture by Eq.~(\ref{eq:spin_enantiomer1}) and Eq.~(\ref{eq:spin_timereversal1}) are also valid for the orbital moment texture. We indeed confirmed this expectation by first-principles calculations [SM Fig. S11]. 
Similarly, the current-induced magnetic moment in tBLG can be switched by the change of the chirality~\cite{bahamon_emergent_2020}.

Recent studies suggest that the resonance between the bands of the graphene and the substrate with strong SOC strength such as topological insulators and transition metal dichalcogenides can enhance the SOC in the graphene \cite{rossi_van_2020,jin_proximity-induced_2013,zhang_proximity_2014,gmitra_graphene_2015,alsharari_mass_2016,gmitra_proximity_2017,arora_superconductivity_2020}. Enhancement of SOC can be as high as two-order of magnitude so that it can lift the spin degeneracy and induce the topological phase transition~\cite{alsharari_mass_2016}.
The chiral spin textures in the single-layer and bilayer graphenes are shown in some of those studies~\cite{jin_proximity-induced_2013,zhang_proximity_2014,gmitra_graphene_2015,gmitra_proximity_2017}. 
If the tBLG is encapsulated with the proximity source in a way that the symmetry is preserved, the SOC splitting may be enhanced so that the spin texture is measurable. In this case, the spin-resolved spectroscopes for the momentum space can be used to observe the spin texture~\cite{tusche_spin_2015, rinaldi_ferroelectric_2018} while the scanning tunneling microscopy (STM) in the magnetic field or the spin-polarized STM may reveal the spin texture according to the symmetry operational similarity~\cite{cheong_permutable_2021}.

\section{Conclusion}
In this work, we focused on the structural chirality  in the tBLG,  defined as the L- and R-twist, which corresponds to the clockwise and counter-clockwise twist of the top layer of bilayer graphene. 
By considering the $\pm 5.086^\circ$ twisted systems as an example, we showed that the vortex of the in-plane displacement pattern appears and its helicity is inverted when one considers a different twist chirality. On the other hand, the out-of-plane displacement pattern is not modified.

We also examined the spin texture of the tBLG around the $K$ point. We found exotic alternating vortex-like spin textures, which do not correspond to the usual Rashba or Dresselhaus type spin texture.
To the best of our knowledge, the spin textures found in this study have never been studied before.
Furthermore, we observed that the helicity of the spin texture is inverted by the twist chirality change since the L-tBLG and R-tBLG are enantiomers. We showed that the spin texture around the $K_1$ and $K_2$ points are related by a $180^\circ$ rotation as a consequence of the time-reversal symmetry. 
We found that the small and large angle tBLGs exhibit different spin texture characteristics.

We propose that the origin of these spin textures can be interpreted in terms of a Rasbha effect in each layer due to the presence of the other layer. In particular, the spin texture can be interpreted as a result of the combined effects of spin textures defined for each layer. In this decomposition, 
we propose model functions based on regression plot models which are able to reproduce the details of the spin textures. 
This decomposition in terms of contributions of spin-texture defined for each model system, 
suggests a possible way to generate exotic spin textures in van der Waals 2D materials by twisting.
It appears that van der Waals heterostructure made of two or more twisted bilayers with different chirality can give rise to interesting effects in terms of electronic and opto-electronic properties, \textit{i.e.}, heterostructure chirality engineering. We believe that this deserves further studies.

\section{Acknowledgement}
A.S. would like to thank Feng-Ren Fan for useful discussions.
J.Y. acknowledges the support by the National Research Foundation of Korea (NRF) grant (No. 2020R1F1A1066548). Additional financial support from Samsung Electronics is also acknowledged.
K.Y. thanks the hospitality by CNR-SPIN c/o Department of Physical and Chemical Science at University of L'Aquila (Italy) during the visiting period from 07/01/2020 to 31/03/2020.
A.S. and S.W.C. acknowledge the CNR Short Term Mobility  program Prot. AMMCNT - CNR n. 80602 dated 18/11/2019.

\bibliography{Spintexture_tBLG}

\pagebreak
\widetext
\begin{center}
\textbf{\large Supplemental Material For ``Chirality-induced spin texture switching in twisted bilayer graphene''}
\end{center}
\setcounter{equation}{0}
\setcounter{figure}{0}
\setcounter{table}{0}
\setcounter{page}{1}
\setcounter{section}{0}
\makeatletter

\renewcommand{\thefigure}{S\arabic{figure}}
\renewcommand{\thesection}{S\Roman{section}}

\section{Structural Relaxation in other cases}

\begin{figure}[t!]
\centering
\includegraphics[width=0.8\textwidth]{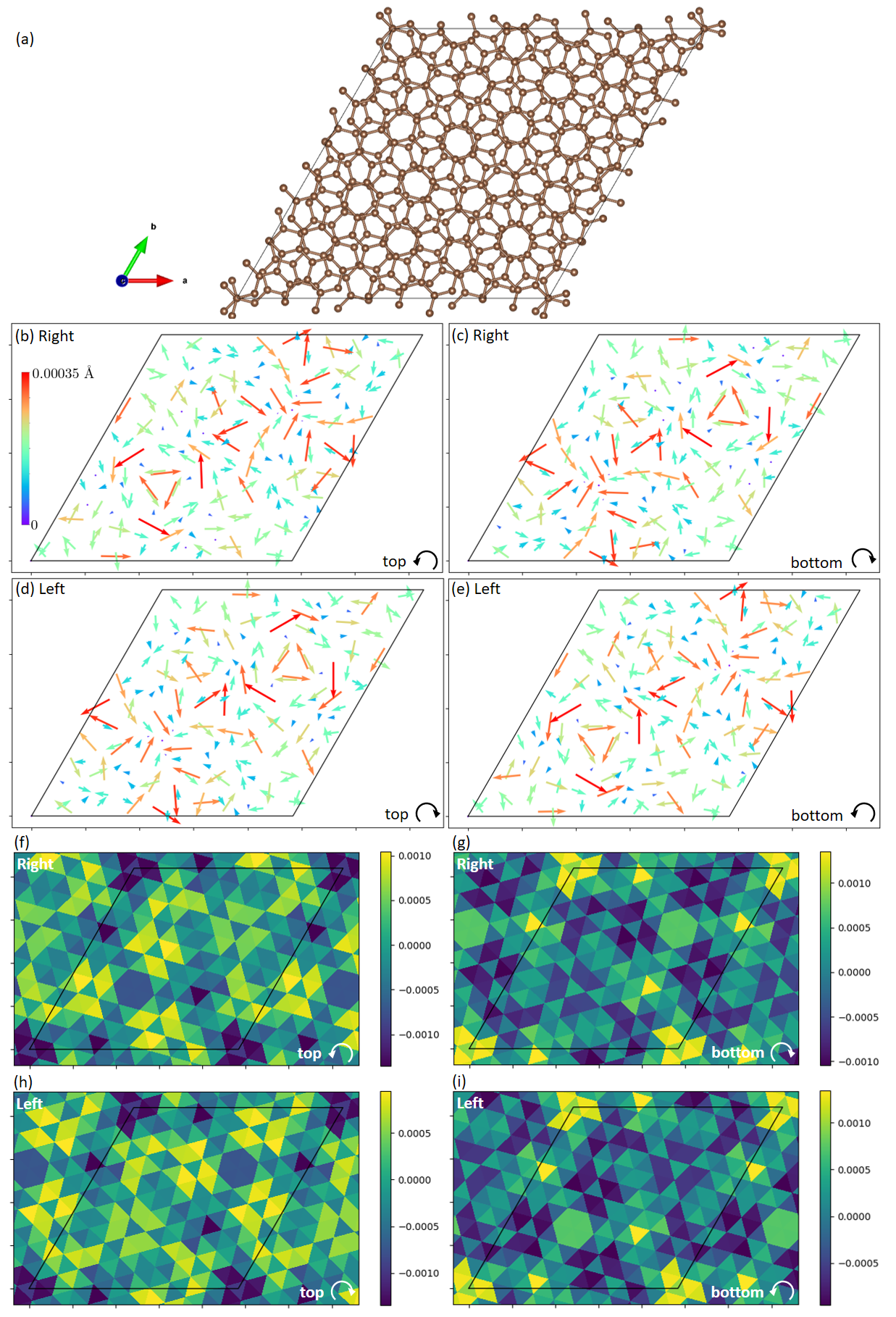}
\caption{
The tBLG with $\pm 29.41^\circ$ twist angle and its displacement patterns.
(a) Supercell.
(b-e) In-plane direction displacements of atoms in the (b) top and (c) bottom layer of tBLG with $+29.41^\circ$ twist (R-chiral). The same displacements for tBLG with $-29.41^\circ$ twist (L-chiral) are shown respectively
in (d) and (e).
(f-i) Color map of the out-of-plane displacements of atoms in (f) top and (g) bottom layer of R-tBLG.
The same quantities are shown respectively in (h) and (i) for L-tBLG.}
\label{fig:system_8_3}
\end{figure}

Let us consider the large twist angle $\pm 29.41^\circ$ tBLG in Fig.~\ref{fig:system_8_3} first. In this case, AA or AB/BA regions are not well-defined as shown in Fig.~\ref{fig:system_8_3} (a). Therefore, the vortex does not clearly appear in the displacement pattern [see Fig.~\ref{fig:system_8_3} (b-e)]. Nevertheless, the fact that the L- and R-tBLG are enantiomers results in the identical displacement pattern of the top layer of R-tBLG and the bottom layer of L-tBLG, and the bottom layer of R-tBLG and the top layer of L-tBLG.
In the out-of-plane displacement pattern Fig.~\ref{fig:system_8_3} (f-i), the positions of the `hill' and `valley' are 
different from those obtained for the small twist angle structures. 
One can see that the maximum magnitude of the displacement is 
much smaller than that obtained in the small twist angle case for both the in-plane and the out-of-plane displacement. Specifically, for the in-plane displacement, 0.0006 \AA\ [small angle, main text Fig. 2] vs 0.00035 \AA\ [large angle, Fig.~\ref{fig:system_8_3} (b-e)], and for the out-of-plane displacement, 0.06 \AA\ [small angle, main text Fig. 3] vs 0.001 \AA\ [large angle, Fig.~\ref{fig:system_8_3} (f-i)].

\begin{figure}[t!]
\centering
\includegraphics[width=0.8\textwidth]{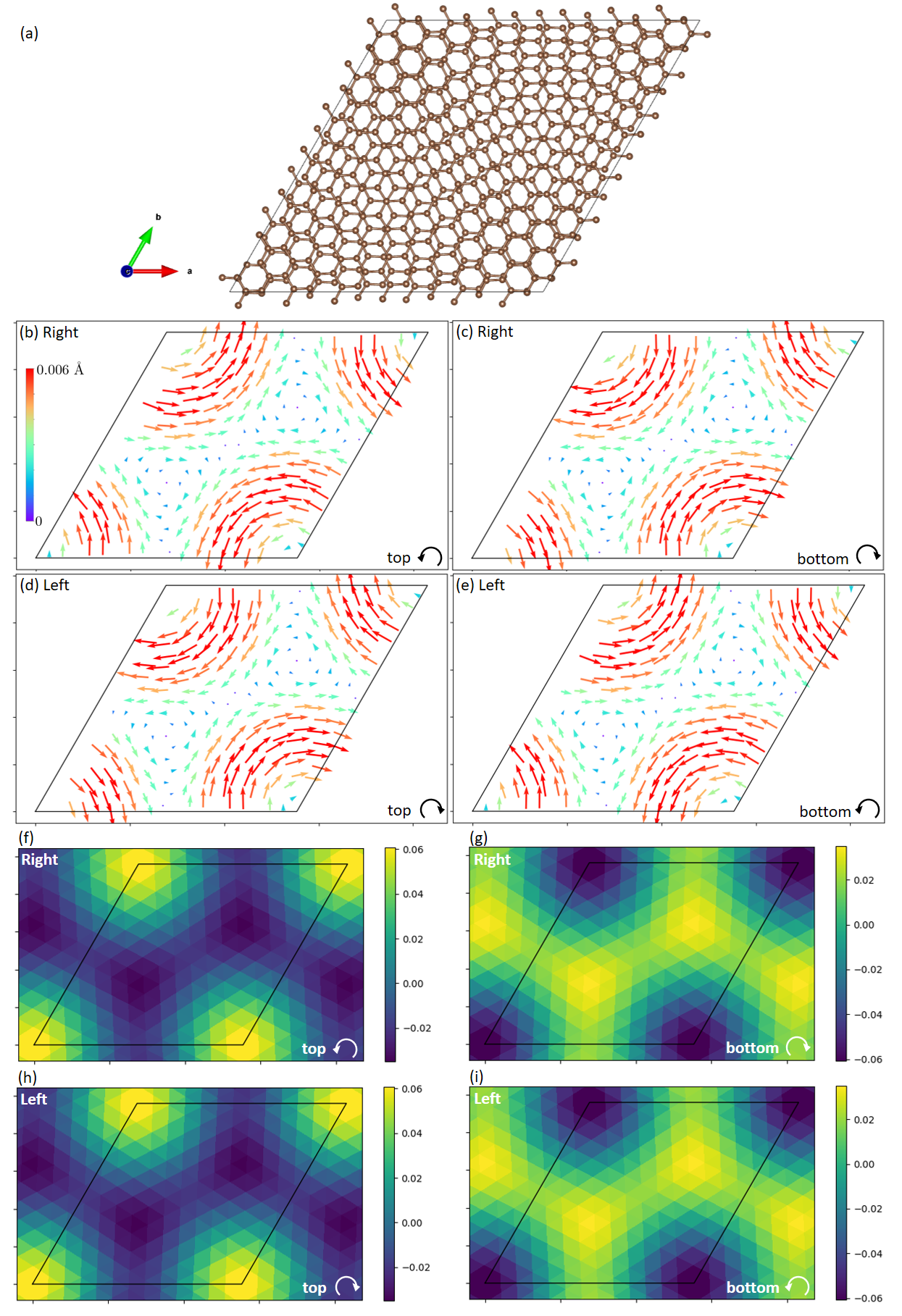}
\caption{
The tBLG with $\pm 5.086^\circ$ twist angle constructed with another symmetry setting, $P622$ space group (no.177) and $D_6$ point group, and its displacement patterns.
(a) Supercell.
(b-e) In-plane direction displacements of atoms in the (b) top and (c) bottom layer of tBLG with $+5.086^\circ$ twist (R-chiral). The same displacements for tBLG with $-5.086^\circ$ twist (L-chiral) are shown respectively
in (d) and (e).
(f-i) Color map of the out-of-plane displacements of atoms in (f) top and (g) bottom layer of R-tBLG.
The same quantities are shown respectively in (h) and (i) for L-tBLG.}
\label{fig:C6_system}
\end{figure}

The case of the tBLG with the $P622$ space group (no.177) and $D_6$ point group corresponding to a different convention for the structures discussed in the main text
is shown in Fig.~\ref{fig:C6_system}. The twist angle is the same as the main text, $\pm 5.086^\circ$.
AA and AB/BA regions are well defined. Moreover, in-plane and out-of-plane displacement patterns show exactly the same behavior and the same magnitude ranges with the $P321$ case in the main text.

\section{Spin textures in other cases}

\begin{figure}[t!]
\centering
\includegraphics[width=0.65\textwidth]{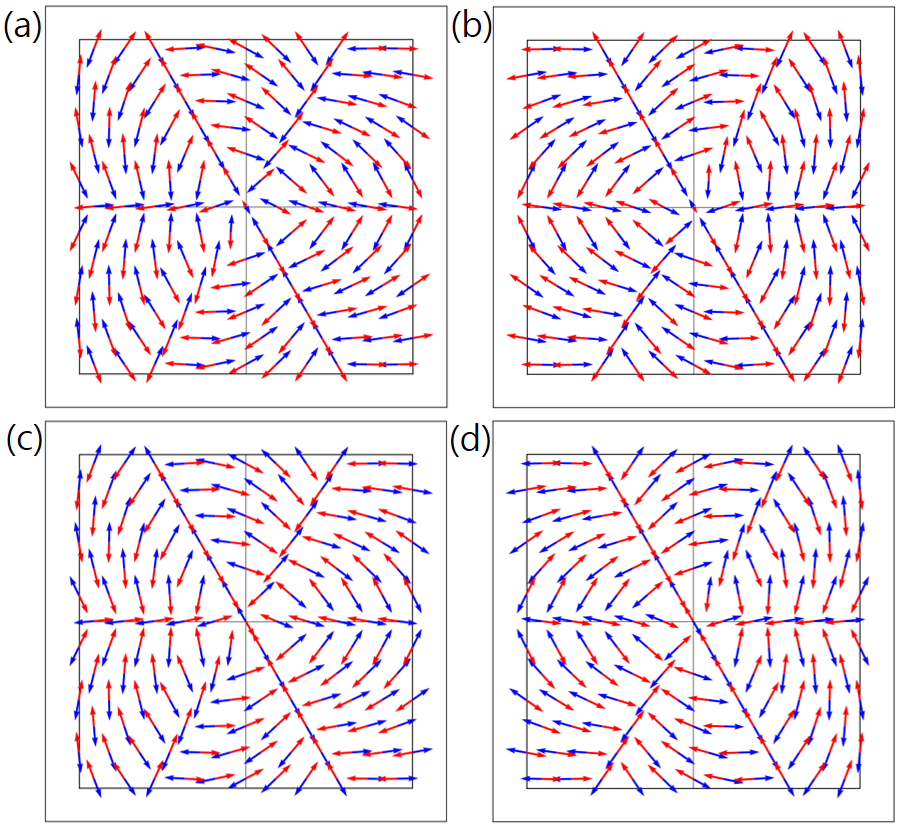}
\caption{
Spin texture of tBLG with $+5.086^\circ$ twist (R-chiral) at (a) $K_1$ and (b) $K_2$ points.
The same quantity for tBLG with $-5.086^\circ$ twist (L-chiral) is shown in (c)  and (d).
The blue arrows are spins of the highest valence band and the red arrows are spins of the second-highest valence band which is a spin split pair of the highest one by the SOC.
}
\label{fig:spin_texture_7_6_upper_lower}
\end{figure}

\begin{figure}[t!]
\centering
\includegraphics[width=0.95\textwidth]{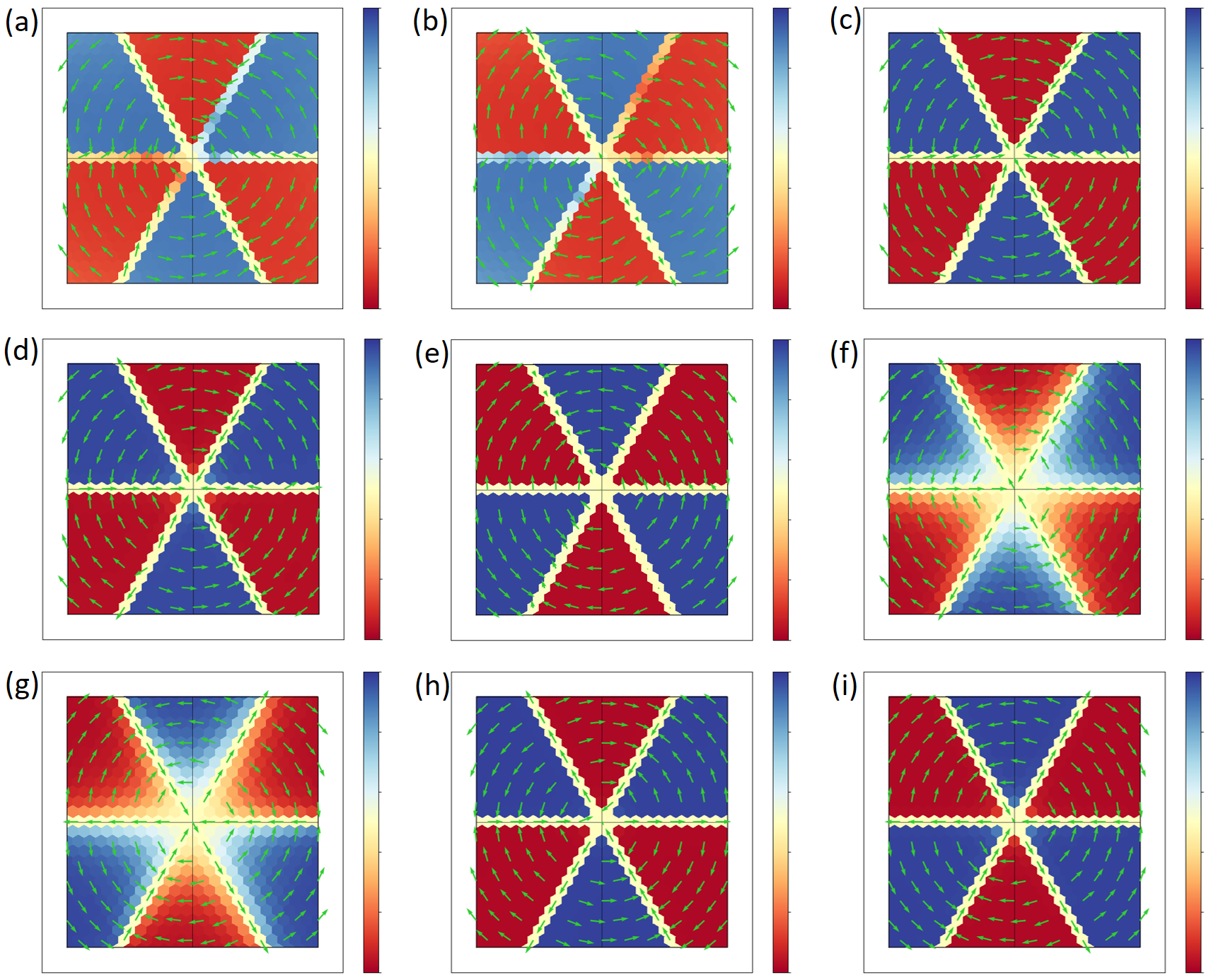}
\caption{
Spin textures and the bottom layer weight distributions of the lowest conduction band around $K_1$ point.
(a) $+5.086^\circ$, (b) $-5.086^\circ$, (c) $+9.430^\circ$, 
(d) $+13.17^\circ$, (e) $+16.43^\circ$, (f) $+21.79^\circ$,
(g) $-21.79^\circ$, (h) $+26.01^\circ$, and (i) $+29.41^\circ$ twist angles.}
\label{fig:conductionband}
\end{figure}

In the main text, only the highest valence band is shown for each system.
Due to the small SOC strength of the carbon atoms, the size of the energy splitting between the upper and lower spin split bands is also very small. The highest and the second-highest valence bands are the upper and lower spin split bands pair. At a given $\mathbf{k}$, the spins of the upper and lower spin split bands have exactly opposite directions with the same size as shown in Fig.~\ref{fig:spin_texture_7_6_upper_lower}.
On the other hand, the spin textures and the layer weight distributions of the lowest conduction band for each system are shown in Fig.~\ref{fig:conductionband}. Although they show similar characteristics with the valence bands, the deviations from the symmetry are larger than the valence bands in $\pm 5.086^\circ$ tBLGs.

\begin{figure}[t!]
\centering
\includegraphics[width=0.8\textwidth]{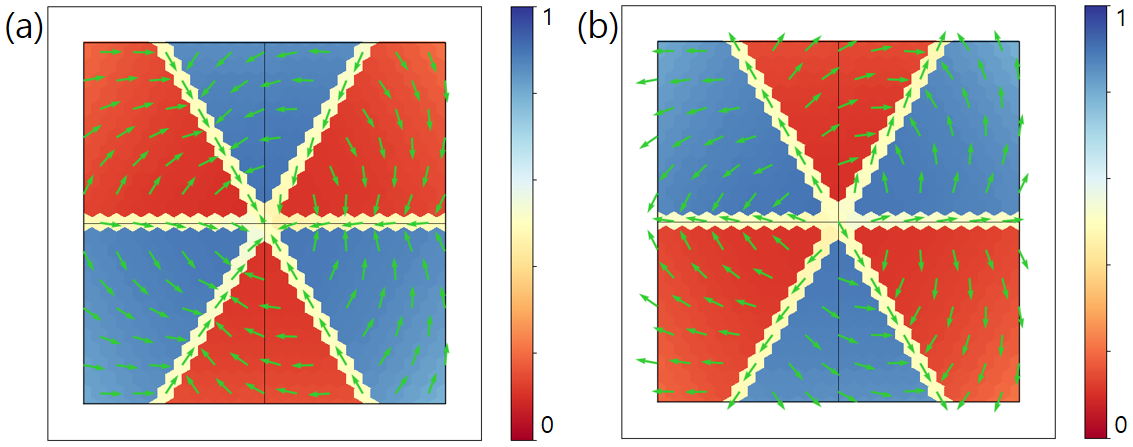}
\caption{
Spin textures and the bottom layer weight distributions around the $K_2$ point of the (a) $+5.086^\circ$ and (b) $-5.086^\circ$ tBLG.}
\label{fig:K2}
\end{figure}

\begin{figure}[t!]
\centering
\includegraphics[width=0.8\textwidth]{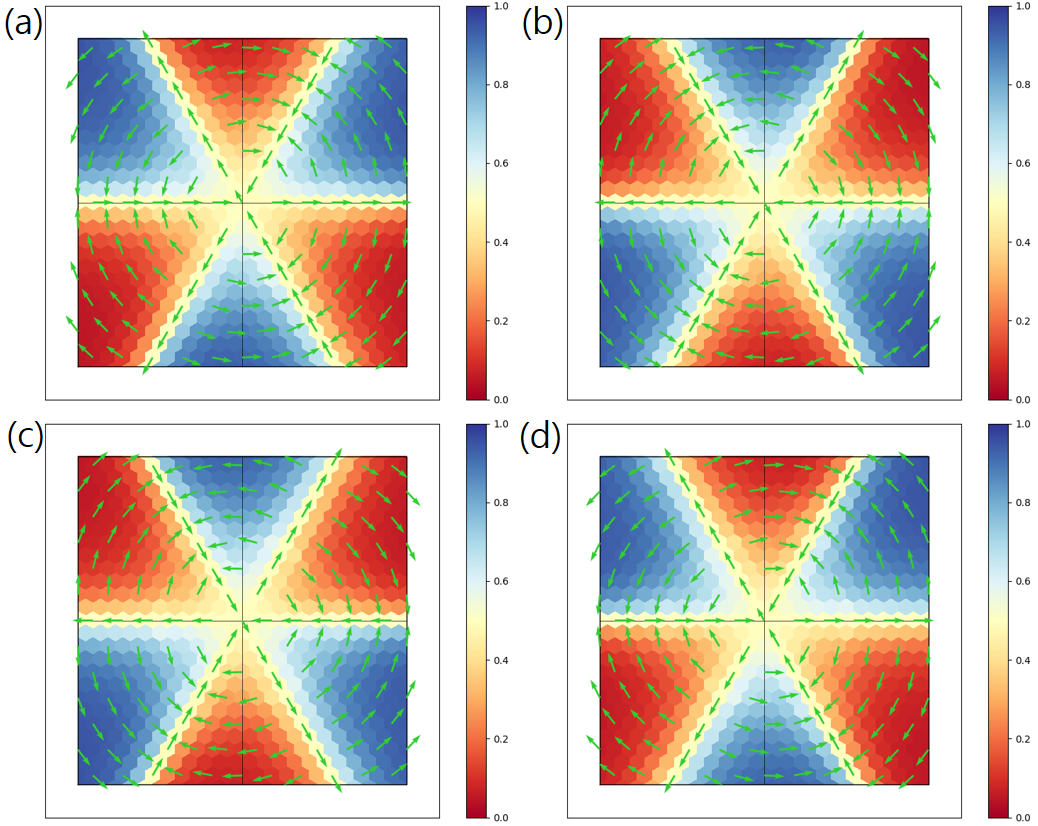}
\caption{
Spin textures of R-tBLG with $+21.79^\circ$ twist at (a) $K_1$ and (b) $K_2$ points.
The same quantities for L-tBLG with $-21.79^\circ$ twist are shown in (c)  and (d).}
\label{fig:spin_texture_2_1}
\end{figure}

\begin{figure}[t!]
\centering
\includegraphics[width=0.8\textwidth]{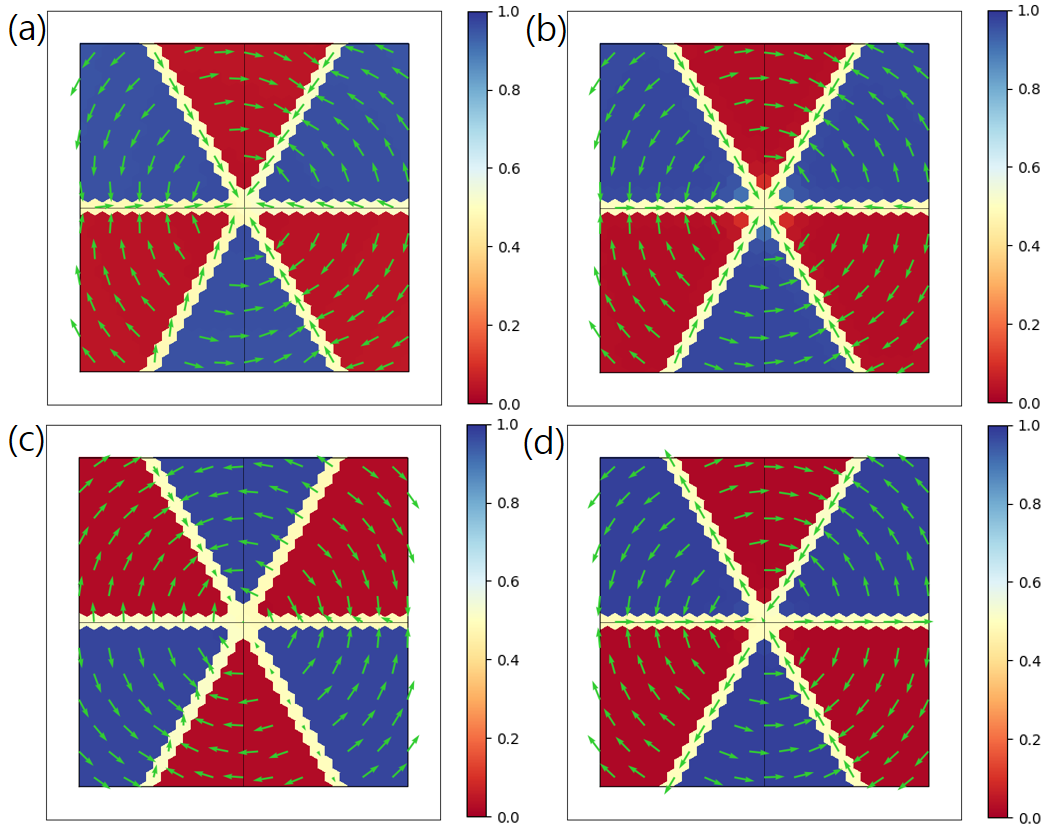}
\caption{
Color maps of the bottom layer weight distribution and the spin textures of the highest valence band around $K_1$ point which are not listed in the main text. 
(a) $+9.430^\circ$, (b) $+13.17^\circ$, (c) $+16.43^\circ$, (d) $+26.01^\circ$ twist angles.}
\label{fig:weights_valence}
\end{figure}

The layer weight distributions around the $K_2$ point of $\pm 5.086^\circ$ tBLG are shown in Fig.~\ref{fig:K2}. It turns out that the switching rule of the layer weight between $K_1$ and $K_2$ point is $180^\circ$ rotation around the $K$ points which is consistent with the switching rule of the spin texture.
Fig.~\ref{fig:spin_texture_2_1} shows the spin texture and layer weight distribution of the $\pm 21.79^\circ$ tBLGs. The figure supports the same switching rule of spin texture and layer weight distribution as the $\pm 5.086^\circ$ tBLGs.
Other layer weight distributions of the highest valence bands for different twist 
angles which are not shown in Fig. 6 of the main text are represented in Fig.~\ref{fig:weights_valence}. All of them also support our arguments relating the spin texture and the layer weight distribution.

\begin{figure}[t!]
\centering
\includegraphics[width=0.8\textwidth]{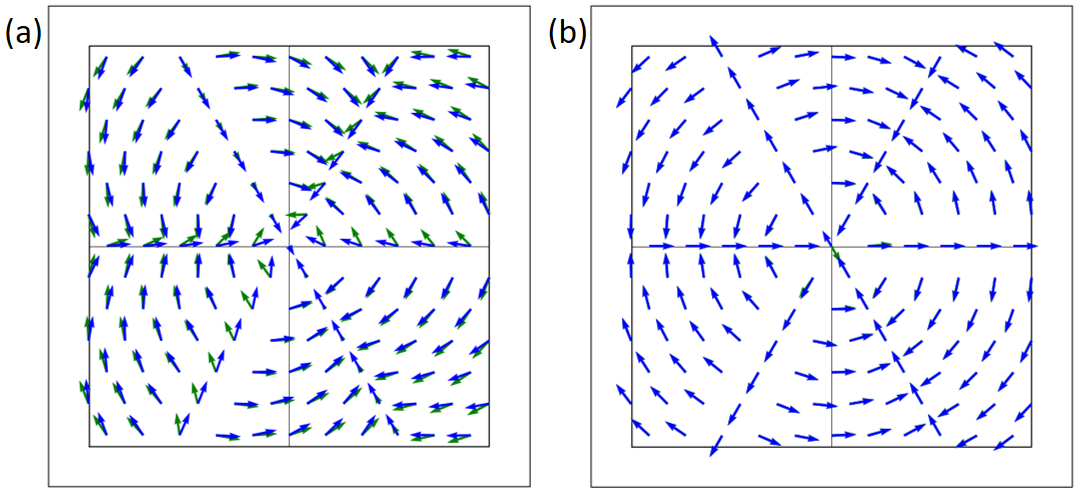}
\caption{
Effect of the structural relaxation on the spin texture of the highest valence band around the $K_1$ point of (a) $+5.086^\circ$ and (b) $+21.79^\circ$ tBLGs.
Blue arrows are the spin textures of the relaxed system and green arrows are the spin textures of the unrelaxed system. }
\label{fig:relaxation_effect}
\end{figure}

The structural relaxation affects the potential and, consequently, the spin texture.
Fig.~\ref{fig:relaxation_effect} shows the comparison of the spin texture before/after atomic relaxations in the $+5.086^\circ$ and $+21.79^\circ$ tBLGs. The blue arrows represent the spin texture after the relaxation which is shown in the main text [Fig. 4 (c) and Fig. 5 (d)] and the green arrows that of the unrelaxed system. 
In the small angle $+5.086^\circ$ case, canting angles of the spins slightly increase after the relaxation. On the 2-fold axes, deviation of the spin from the symmetry line is reduced as a result of the relaxation. Spins in the large angle $+21.79^\circ$ case almost remain unchanged after the relaxation. This is consistent with the small displacements after relaxations in the large angle tBLGs.

\section{Model plot for spin textures in other cases}

\begin{figure}[t!]
\centering
\includegraphics[width=0.8\textwidth]{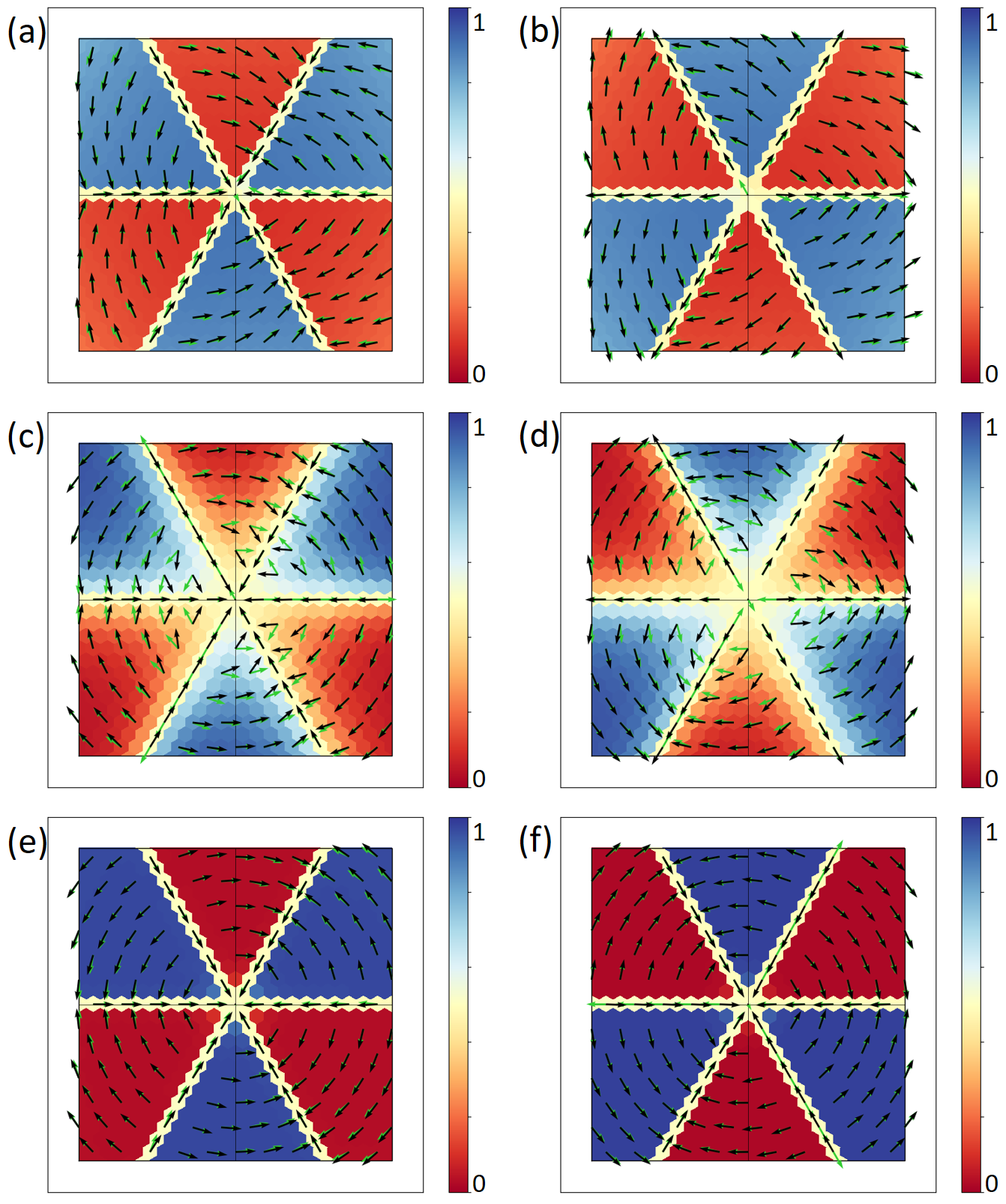}
\caption{
Model spin texture from the canting model 1 (black arrows), calculated spin texture (green arrows), and bottom layer weight distribution (color map) including the cases which are not listed in the main text.
(a) $+5.086^\circ$, (b) $-5.086^\circ$, (c) $+21.79^\circ$, (d) $-21.79^\circ$, (e) $+13.17^\circ$, (f) $+29.41^\circ$ twist angles.}
\label{fig:cantingmodel}
\end{figure}

\begin{figure}[t!]
\centering
\includegraphics[width=0.8\textwidth]{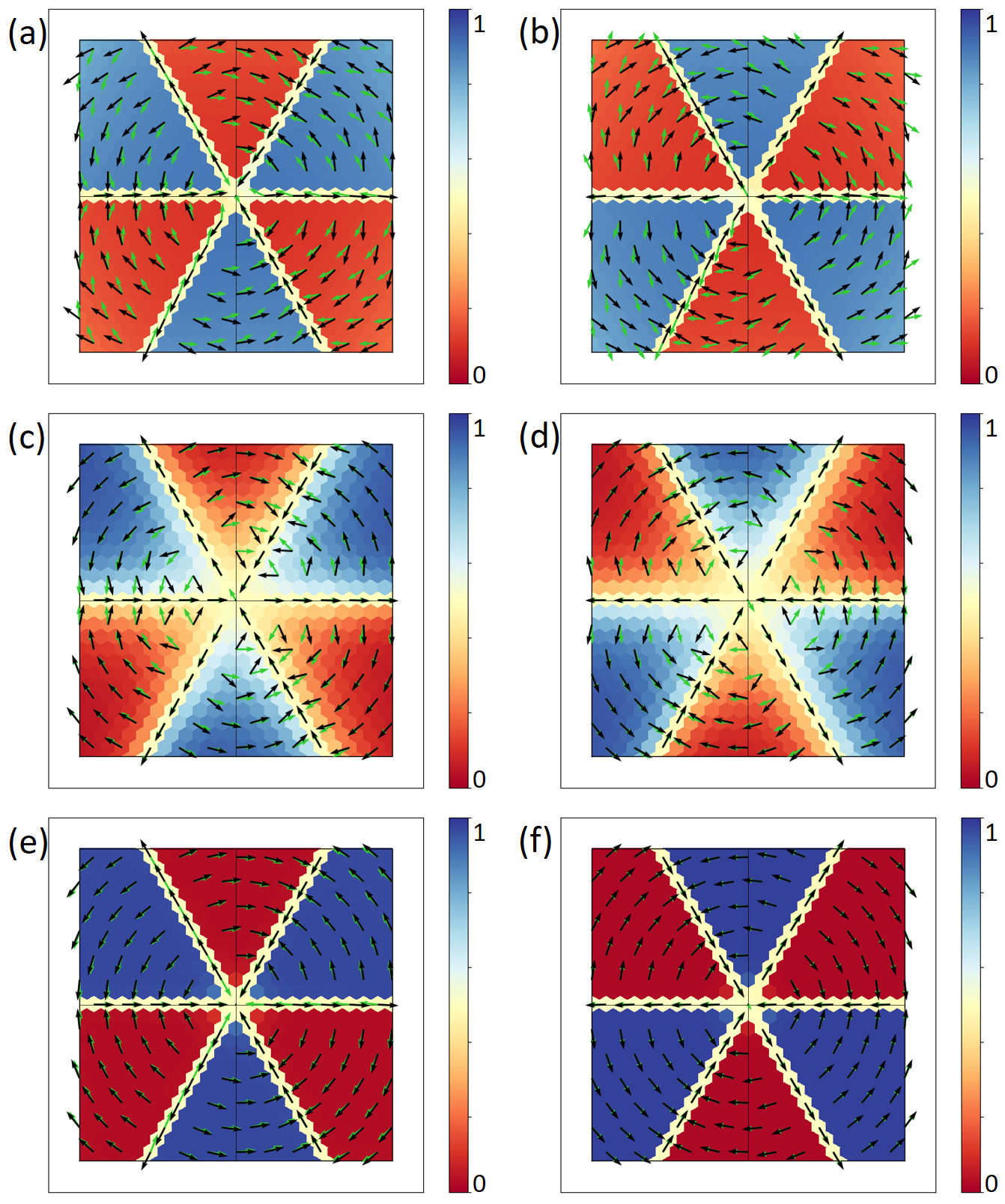}
\caption{
Model spin texture from the trigonal warping model 2 (black arrows), calculated spin texture (green arrows), and bottom layer weight distribution (color map) including the cases which are not listed in the main text.
(a) $+5.086^\circ$, (b) $-5.086^\circ$, (c) $+21.79^\circ$, (d) $-21.79^\circ$, (e) $+13.17^\circ$, (f) $+29.41^\circ$ twist angles.}
\label{fig:TWmodel}
\end{figure}

The model plots for some cases that are not listed in the main text are shown in Fig.~\ref{fig:cantingmodel} for the canting model 1 and Fig.~\ref{fig:TWmodel} for the trigonal warping model 2. For a better comparison, plots that are already shown in the main text are included again.

\section{Orbital angular momentum}

\begin{figure}[t!]
\centering
\includegraphics[width=0.8\textwidth]{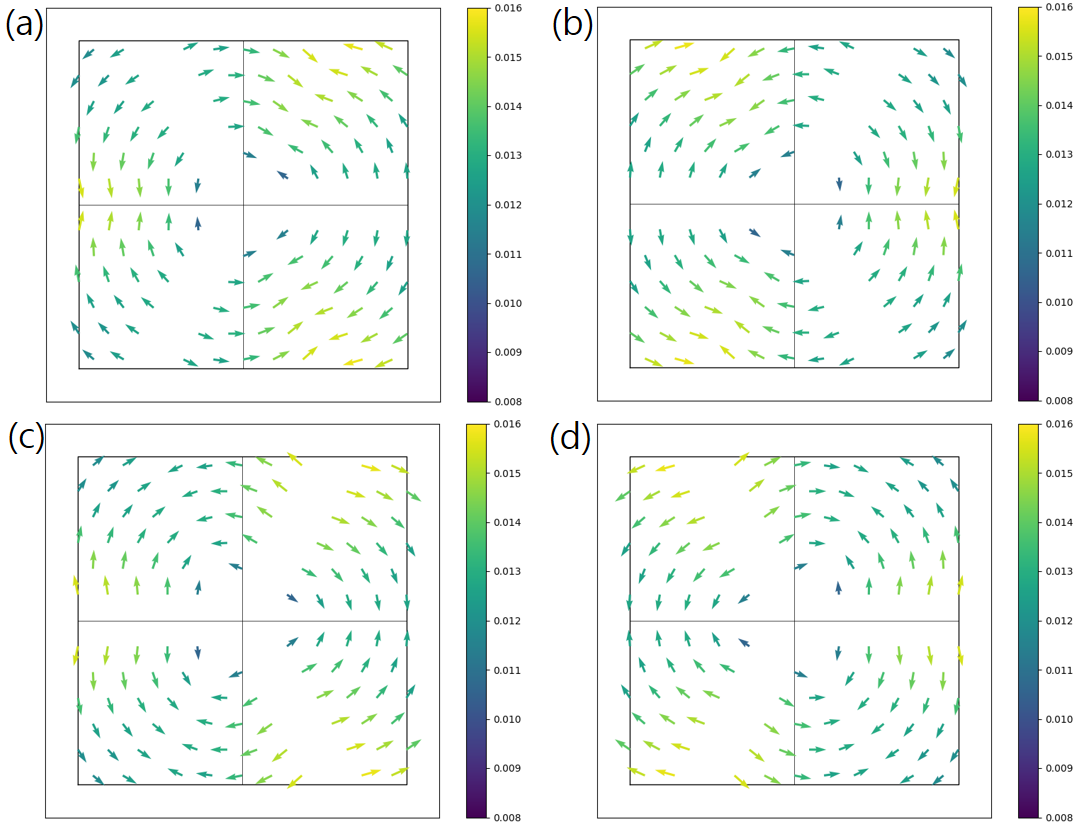}
\caption{
Orbital angular momentum texture of tBLG with R-chiral tBLG with $+21.79^\circ$ twist at (a) $K_1$ and (b) $K_2$ points.
The same quantity for tBLG with L-chiral $-21.79^\circ$ twist is shown in (c)  and (d).
Color map represents the magnitude of the orbital angular momentum in $\hbar$ units.
}
\label{fig:orbital_texture_2_1}
\end{figure}

The orbital angular momentum texture follows the same switching rules discussed in the main text. Fig.~\ref{fig:orbital_texture_2_1} shows the orbital angular momentum of the highest valence band of the $\pm 21.79^\circ$ tBLGs around the $K$ points. 
Orbital angular momentum is calculated by using another DFT program OpenMX [S1].
OpenMX is a DFT code adopting the Linear Combination of the Pseudo Atomic Orbital (LCPAO) method. We take two $s$ orbitals and two sets of $p$ orbitals $\{p_x,p_y,p_z\}$ optimized for the carbon atoms as the basis set which is denoted as \texttt{C6.0-s2p2} in the OpenMX. The GGA-PBE exchange-correlation functional is used [S2]. The orbital texture is not shown along the 2-fold axes for simplicity.

\section{Brillouin zone folding patterns}

\begin{figure}[t!]
\centering
\includegraphics[width=0.8\textwidth]{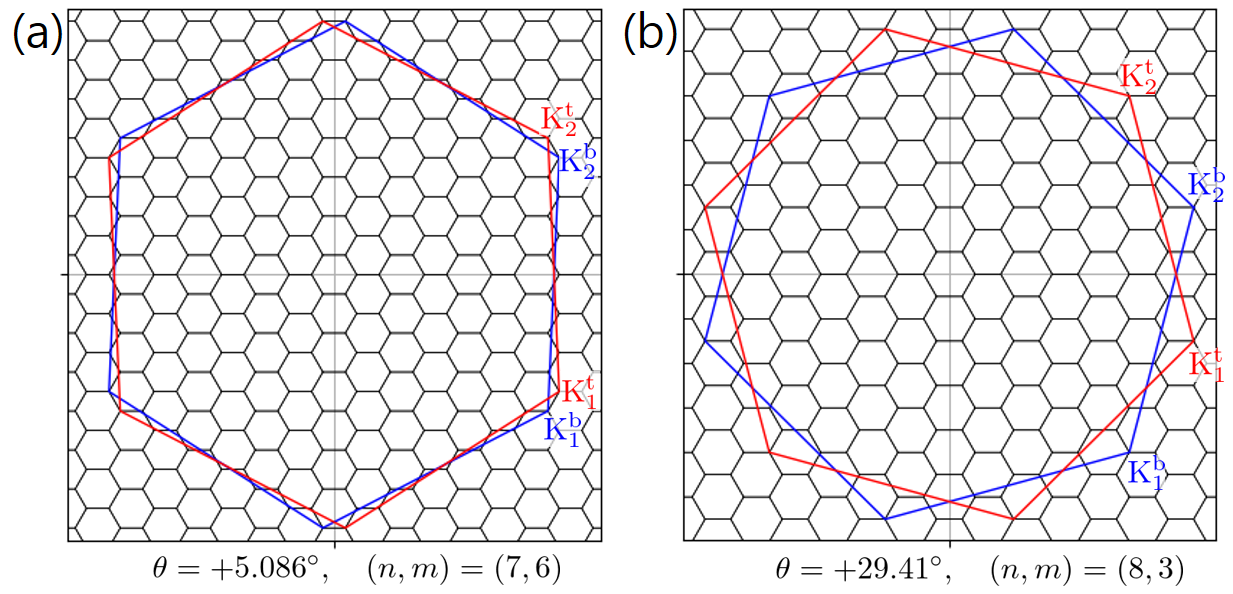}
\caption{
Different Brillouin zone (BZ) folding pattern into SBZ between the (a) $+5.086^\circ$ tBLG with $(n,m)=(7,6)$ and (b) $+29.41^\circ$ tBLG with $(n,m)=(8,3)$.
The red line represents the boundary of unfolded BZ of the top layer and the blue one represents that of the bottom layer. The black line represents the SBZ and its periodic copies. }
\label{fig:BZfolding}
\end{figure}

In spin textures shown in Fig$.\ 5$ of the main text, one can see that the (c) $\theta=+16.43^\circ$ and (f) $\theta=+29.41^\circ$ cases have the opposite helicities to those of other cases (a,b,d, and e).
That is because the folding patterns of the Brillouin zone (BZ) of the top and bottom layer into the SBZ are different for those twist angles. This is shown in Fig.~\ref{fig:BZfolding} representing the BZ folding of the $+5.086^\circ$ tBLG [panel (a)] and $+29.41^\circ$ tBLG [panel (b)].
In the $+5.086^\circ$ tBLG of (a), $K_1^\text{t}$, \textit{i.e.} the $K_1$ point of the unfolded BZ of the top layer, corresponds to the $K_1$ point of one of the periodic copies of the SBZ. 
It means that $K_1^\text{t}$ is folded into $K_1$ point of the SBZ. 
On the other hand, $K_1^\text{b}$ of the bottom layer corresponds to the $K_2$ point of the SBZ. $K_2^\text{t}$ is at $K_2$ and $K_2^\text{b}$ is at $K_1$.
However, in the $+29.41^\circ$ tBLG of (b), $K_2^\text{t}$ and $K_1^\text{b}$ ($K_1^\text{t}$ and $K_2^\text{b}$) are placed at $K_1$ ($K_2$).

The helicity change is explained as follows. At the $K_1$ point, for example, ($K_1^\text{t}$, $K_2^\text{b}$) pair at $+5.086^\circ$ is replaced with ($K_1^\text{b}$, $K_2^\text{t}$) pair at $+29.41^\circ$, \textit{i.e.}, the top and bottom are exchanged with each other. As a consequence, the dominantly weighted layers are also exchanged. Finally, according to our layer weight interpretation for the spin texture, also the helicities are exchanged.
Among the angles in Table \MakeUppercase{\romannumeral 1} of the main text, $\theta=+16.43^\circ$ and $\theta=+29.41^\circ$ cases correspond to Fig.~\ref{fig:BZfolding} (b), and other cases correspond to (a). 
One can find after a little algebra that $n-m=1$ is the condition 
that $\overline{K_2^\text{t}K_2^\text{b}}$ defines a side of the SBZ. 
This condition always results in the BZ folding pattern corresponding 
to (a) and makes the SBZ equal to the MBZ in the small angle cases. 

\hfill

\hfill

\textbf{References for SM}

\hfill

[S1] T. Ozaki, Phys. Rev. B \textbf{67}, 155108 (2003), website : openmx-square.org

[S2] J. P. Perdew, K. Burke, and M. Ernzerhof, Phys. Rev. Lett. \textbf{77}, 3865 (1996)

\end{document}